\documentclass[final,3p,times,twocolumn,authoryear]{elsarticle}

\usepackage{epsfig}
\usepackage{adjustbox}

\usepackage{amssymb}
\usepackage{amsmath}
\usepackage{amsthm}

\usepackage{lineno}
\usepackage{xcolor}
\usepackage{ulem}
\usepackage{float}
\usepackage{bm}

\usepackage{natbib}
\usepackage{url} 

\usepackage{multirow}

\newcommand*{\geant}{GEANT4}

\journal{Nuclear instrumentation and methods in physics research, A}

\begin{document}
\begin{frontmatter}



\title{Prediction of the SVOM MXT camera end of life spectral performance based on proton irradiation results}


\author[inst1]{Clara Plasse}

\affiliation[inst1]{organization={AIM-CEA/DRF/Irfu/Departement d’Astrophysique, CNRS, Universite Paris-Saclay, Universite Paris Cite},
            addressline={Orme des Merisiers, Bat. 709}, 
            city={Gif-sur-Yvette},
            postcode={91191}, 
            country={France}}

\author[inst1]{Diego Götz}
\author[inst1]{Aline Meuris}
\author[inst1]{Miguel Fernandez Moita}
\author[inst1]{Philippe Ferrando}

\author[inst2]{Leo Favier}

\author[inst3]{Francesco Ceraudo}

\affiliation[inst2]{organization={Nucletudes, AIM-CEA/IRFU/Departement d’Astrophysique, CNRS, Universite Paris-Saclay, Universite Paris Cite},
            addressline={3 avenue du Hoggar},
            city={Les Ulis},
            postcode={91940}, 
            country={France}}
            
\affiliation[inst3]{organization={INAF/IAPS},
            addressline={Via del Fosso del Cavaliere 100}, 
            city={Roma},
            postcode={00133}, 
            country={Italy}}

\begin{abstract}
SVOM, the Space-based Variable astronomical Object Monitor, launched on June 22$^{nd}$ 2024, is a Chinese-French mission focused on exploring the brightest phenomena in the cosmos – Gamma-Ray Bursts. Among the four instruments on board is the Micro-channel X-ray Telescope (MXT). The MXT camera features a 256 $\times$ 256 pixel pnCCD detector to perform X-ray imaging and spectroscopy in the 0.2--10 keV energy range. Cruising in a low-Earth orbit ($\sim$ 600 km) that crosses the South Atlantic Anomaly, the MXT focal plane is exposed to radiation, primarily protons, that will lead to performance degradation over time. The challenge for MXT, and possibly for future missions with similar mass and mechanical constraints, is to maintain spectral performance all along the mission duration. To assess the expected radiation-induced performance degradation, a spare flight model of MXT focal plane underwent an irradiation campaign with 50 MeV protons at the Arronax cyclotron facility in June 2022. Then, the proton irradiated spare model was characterized in detail at the X-ray Metrology beamline of the SOLEIL Synchrotron facility in June 2023, as well as with a laboratory X-ray fluorescence source. We find through the evaluation of key indicators of performance such as the charge transfer inefficiency (CTI) and the low energy threshold, that MXT will remain compliant to its requirements over the SVOM mission lifetime. We also report an unexpected effect of proton irradiation that is the inversion of the trend of CTI with energy, recovered with two different sources illuminating the detector, and never reported in literature so far.

\end{abstract}


\end{frontmatter}


\section{Introduction}
\label{sect:intro}
SVOM (Space-based Variable astronomical Object Monitor) is a space mission aimed at studying Gamma-Ray Bursts (GRBs) and other rapid, high-energy transient events across multiple wavelengths (\cite{SVOM}, \cite{SVOM_2}). It was launched in June 2024 on a circular low Earth orbit (h~$\sim$~600 km) with an inclination of about 29$^{\circ}$. Along its low-Earth orbit, due to South Atlantic Anomaly crossings, the SVOM satellite and its instruments is subject to radiation damage, primarily from proton interactions. In the case of The Micro-channel X-ray Telescope (MXT), one of the four main instrument aboard, the proton irradiation creates defects in the detector crystalline structure, expected to cause performance degradation. 

\begin{figure}
\begin{center}
\includegraphics[height=5cm]{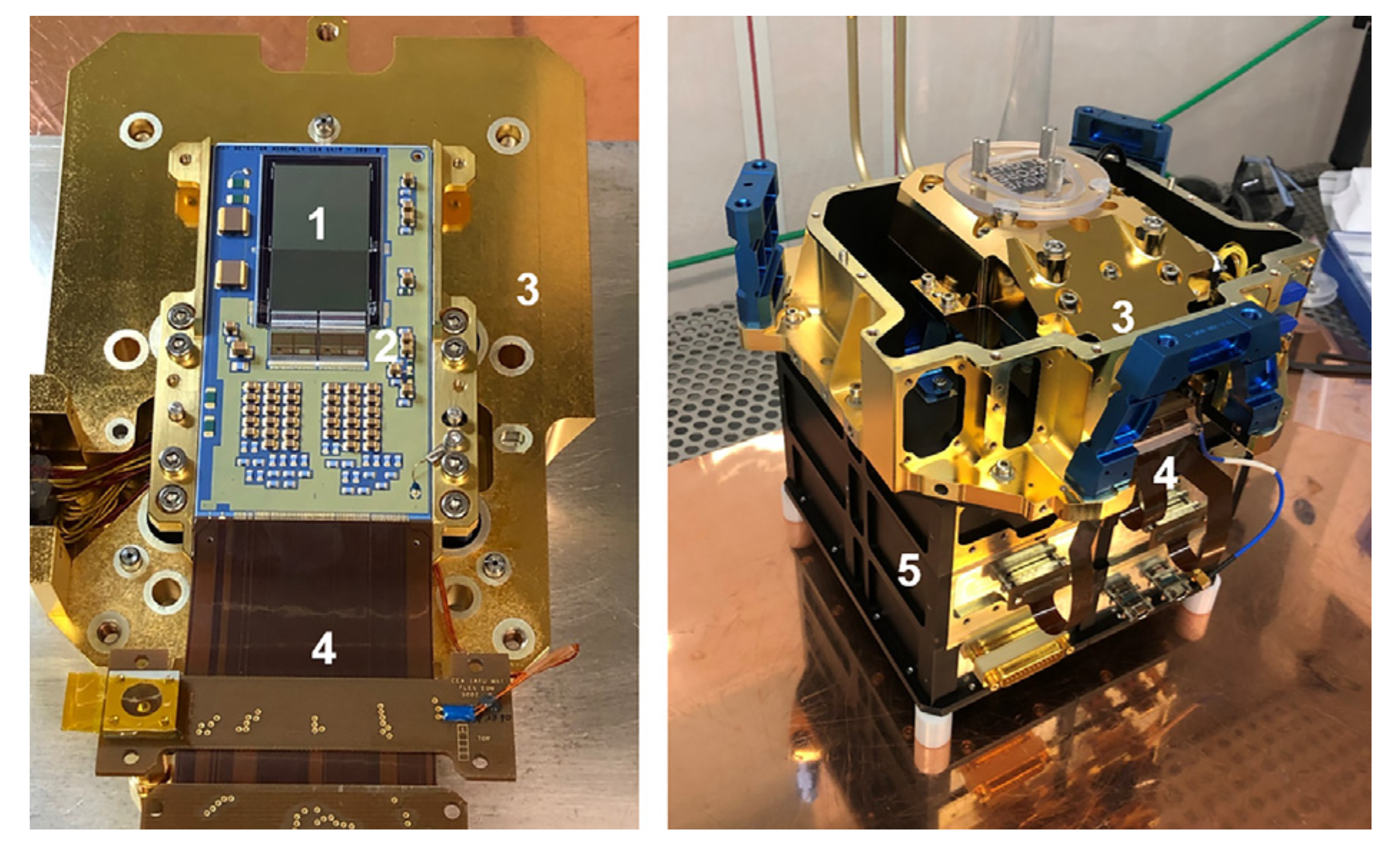}
\end{center}
\caption 
{ \label{fig:mcam}
Focal plane assembly (FPA) of MXT camera. A ceramic board equipped with a pnCCD (1) and 2 Camex ASICs (2) is mounted in a proton shielding (3) and the flex lead (4) is connected to the warm front-end electronics (5).} 
\end{figure}

The MXT camera is composed of the Focal Plane Assembly (FPA), a Front-End Electronics (FEE) assembly and a Calibration Wheel Assembly to perform X-ray imaging and spectroscopy in the 0.2--10 keV energy range, as illustrated in Figure \ref{fig:mcam}. The calibration wheel features four positions: (i) two open position for sky observations, (ii) a calibration position utilizing a radioactive $^{55}$Fe source that fully illuminates the detector, and (iii) a closed position equipped with a 10~mm thick copper shutter for regular transitions through the South Atlantic Anomaly. The core of the camera is a pnCCD (450 $\mu$m-thick fully depleted charge coupled detector made of pn junctions using the sidewards depletion principle) with a 256 $\times$ 256 pixel imaging area for photon integration and a frame store area for the image readout. This imaging detector, optimized for high resolution X-ray spectroscopy, was designed, produced and first characterized by the Max-Planck-Institute for Extraterrestrian Physics (MPE) in the context of the DUO mission \citep{duo}. It has been implemented in the context of the SVOM mission, taking advantage of the last generation of front-end CAMEX ASICs designed and being successfully operated in the eROSITA instrument on board the SRG mission \citep{erosita}, and by developing a dedicated readout electronics system compatible with the exportation rules to China. The operating temperature of $-65^\circ$C of the detector is a compromise between the cooling capability for this small satellite on a low-Earth orbit (an intrinsically highly variable thermal environment) and the necessity to limit the dark current of the detector. Previous missions with the same detector technology operate the pnCCD at significantly colder temperatures: $-90^\circ$C for EPIC/XMM-Newton launched in 1999 \citep{XMM} and for FXT/Einstein Probe launched in 2023 \citep{EP}, and $-80^\circ$C for eROSITA/SRG launched in 2019 \citep{SRG}. The capability to maintain the required spectral performance along the mission lifetime is consequently a serious concern and challenge for SVOM and for future missions with similar orbits, power and mass resources. The target performance is an energy resolution better than 80 eV FWHM at 1.5~keV at beginning of life and better than 200~eV FWHM (+250 ~\%) after the 3 years of nominal mission. 

To monitor the spectral evolution in flight, the calibration strategy relies on the observation of two types of sources: an internal $^{55}$Fe radioactive source assembled to a filter wheel, and astrophysical sources. Astrophysical sources, typically Supernova Remnants (SNRs) are stable in flux and emit lines at numerous energies across the X-ray domain. However, due to the SNRs lower flux (wrt. the internal source), they require longer exposure times, which could potentially conflict with the mission observation schedule. On the contrary, the calibration source can be used regularly, e.g. during each Earth occultations, fully illuminating the detector and emitting two spectral lines around $\sim$6~keV. Given that the maximum sensitivity of MXT is around 1~keV, very precise calibration and a good understanding of the detector response at all energies are crucial to correctly extrapolate the regular $^{55}$Fe measurements over the whole energy range. This will be particularly necessary for an ageing MXT camera, as the spectral response will deteriorate over time and could even affect the instrument's source location capability, if the low-energy threshold increases excessively. For a complete description of the MXT telescope, see \cite{Diego}.

The paper is organized as follows: we describe the methodology for the performance degradation assessment, including the campaigns carried out for the irradiation and characterization of MXT focal plane, and the data analysis techniques employed to forecast performance throughout the mission lifespan in Section \ref{sect:methods}. Our empirical findings and interpretations focused on the effect of irradiation are outlined in Section \ref{sect:results}. Finally, we discuss these results and review the expected and unexpected effects of the proton irradiation in Section \ref{sect:discussions}.

\section{Experimental setup and performance characterisation}
\label{sect:methods}

\subsection{The MXT camera}

The detector board shown in Figure \ref{fig:mcam}, made of aluminium oxide ($Al_2O_3$), holds the pnCCD, the two CAMEX ASICs and passive parts for voltage filtering. It is glued on a copper-molybdenum mechanical structure which protects the frame store area from X-rays and which is thermally coupled to the active cooling system. The main features of MXT camera are listed in Table \ref{tab:MXT_characteristics}.

\begin{table}[ht]
\caption{MXT instrument main characteristics} 
    \label{tab:MXT_characteristics}
    \begin{center}
    \scalebox{0.72}{
    \begin{tabular}{c|c}
    \hline
    \rule[-1ex]{0pt}{3.5ex} Detector type & pnCCD DUO  \\ 
    \hline
    \rule[-1ex]{0pt}{3.5ex} ASIC type & CAMEX 128MJD \\
    \hline
    \rule[-1ex]{0pt}{3.5ex} Image array & 256 x 256 pixels  \\ 
    \hline
    \rule[-1ex]{0pt}{3.5ex} Detector active area & 1.92~$mm^2$ x 1.92~$mm^2$ \\ 
    \hline
    \rule[-1ex]{0pt}{3.5ex} Pixel size & 75~$\mu$m  \\ 
    \hline
    \rule[-1ex]{0pt}{3.5ex} Quantum efficiency & 27\% at 0.28~keV \\ 
    \rule[-1ex]{0pt}{3.5ex}  & 99\% at 4.5~keV \\
    \hline
    \rule[-1ex]{0pt}{3.5ex} (Image area) Integration time & 100~ms  \\ 
    \hline
    \rule[-1ex]{0pt}{3.5ex} (Frame store) Readout time & 8~ms  \\ 
    \hline
    \rule[-1ex]{0pt}{3.5ex} Readout mode & Full frame or Event (main operational mode)  \\ 
    \hline
    \end{tabular}}
    \end{center}
\end{table}

\begin{figure*}[h]
\begin{center}
\begin{tabular}{c}
\includegraphics[height=6cm]{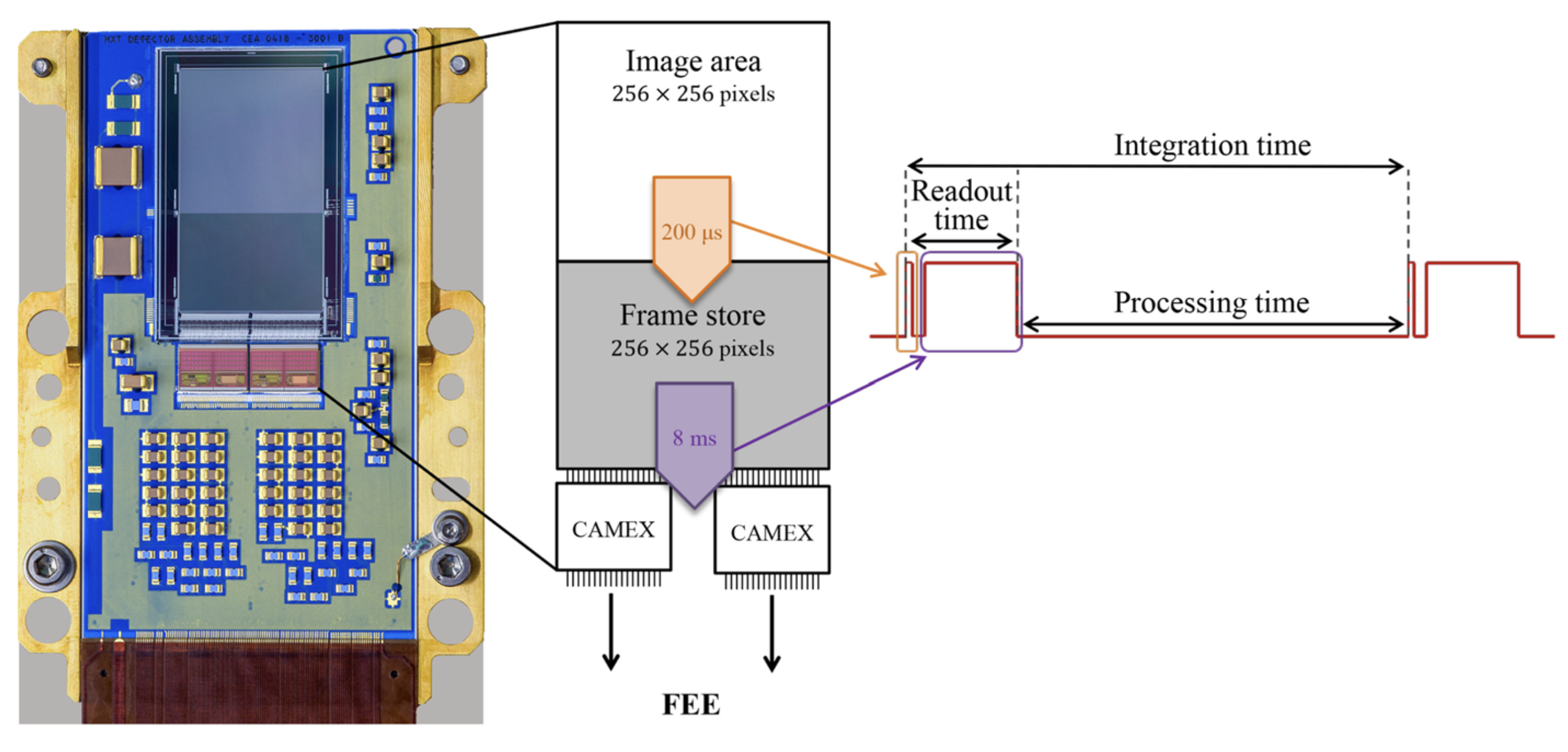}
\end{tabular}
\end{center}
\caption 
{ \label{fig:FPA+FEE}
Representation of MXT FPA. On the right, a schematic of the detector with the typical sequence to collect a frame and process it in the FEE is shown. } 
\end{figure*} 

Photons enter by the planar entrance window of the exposed image area through its backside, and create by photoelectric effect a charge cloud. Those charges then drift towards the front side, and are kept under individual pixel structure (or several neighbor pixels in case of charge sharing, see below). Then the charges are simultaneously transferred to the shielded frame-store area, in about 200~$\mu$s. While the charges are being transferred to the anodes of the pnCCD columns, the frame store read-out is performed row by row. The signals from each column are converted into voltages using the on-chip JFETs, and then amplified and filtered by the CAMEX analog channels. During the processing of the next row, these signals are multiplexed in a timely manner at the CAMEX analog outputs, enabling a frame store readout in only 8 ms. A schematic summary of the detection chain can be found Figure \ref{fig:FPA+FEE}.

At the CAMEX output, the FEE ensures the digitization of the analog channels and carries out mode-dependent processing of the pixel data prior to sending it to the MXT data-processing unit (MDPU). Additionally, the FEE supplies all bias voltages for the CAMEX ASIC and the pnCCD, along with the readout control signals. The data processing depends on the mode MXT camera is operated in. There are two main modes of operation: the \textit{full frame} mode and the \textit{event} mode. \textit{Full frame} mode transmits all data collected by the detector to the MDPU, but at lower rate than the \textit{event} mode (5 frames per second). Due to telemetry limitations, the \textit{full frame} mode is only used for so-called {\it dark measurements} with the wheel in closed position, to obtain a measurement of the offset level and the noise level of every pixel of the detector matrix without photon illumination. For each pixel, the Low Level Threshold (LLT) is defined as a configurable $n$ times the noise value of that pixel. During ground tests, the value of $n=4$ turned out to be a good compromise to ensure the extraction of X-ray events and obtain the best energy resolution. The values of each pixel's LLT are then uploaded to the FEE to be used in the \textit{event} mode. In this mode, the FEE transmits only pixels with signal larger than the LLT value, i.e well above noise fluctuations. The \textit{event} mode is the nominal read-out mode, operated at the rate of 10 frames per second. This means that for each exposure of 100 ms, we get a list of hit pixels that correspond to X-ray events candidates with optimal energy resolution, while still keeping the data sent by the satellite to ground within a manageable amount.

\subsection{Predictions of the displacement damage dose}

In this paper, we focus on the long term effects of the space radiation environment, not on the transient effects (single events) that high energetic particles can cause on the detection chain when it is in operation. In order to accurately estimate the fluence the MXT detector will be subjected to, a GEANT4 Monte Carlo simulation \citep{francesco_these} was developed with the \geant ~toolkit \citep{geant4_1, geant4_2}. Table~\ref{tab:sim:fluence} summarizes the results. 

\begin{table*}[h]
	\caption{Summary of the estimations of the 10 MeV and 50 MeV proton equivalent fluence obtained from \geant ~for four values of the time spent in orbit.}
	\label{tab:sim:fluence}
	\centering
	\begin{tabular}{lllll} 
	\hline
	{}  & \multicolumn{1}{c}{$1\,\mathrm{yr}$} & \multicolumn{1}{c}{$3\,\mathrm{yr}$} & \multicolumn{1}{c}{$5\,\mathrm{yr}$} & \multicolumn{1}{c}{$10\,\mathrm{yr}$} \\
	{} & \multicolumn{1}{c}{$\mathrm{cm}^{-2}$} & \multicolumn{1}{c}{$\mathrm{cm}^{-2}$} & \multicolumn{1}{c}{$\mathrm{cm}^{-2}$} & \multicolumn{1}{c}{$\mathrm{cm}^{-2}$} \\
	\hline
	{} 10 MeV equivalent fluence & $8.24\times 10^{8}$ & $2.47\times 10^{9}$ & $4.12\times 10^{9}$ &	$8.24\times 10^{9}$  \\
	{} 50 MeV equivalent fluence  & $2.03\times 10^{9}$ & $6.09\times 10^{9}$ & $1.02\times 10^{10}$ & $2.03\times 10^{10}$ \\	
	\hline
	\end{tabular}
\end{table*}

In the GEANT4 model, the detector was placed inside a simplified version of the whole scientific payload of the SVOM spacecraft, and at the center of an isotropic particle distribution consisting of protons with a spectrum equal to the average spectrum of protons trapped in Earth's magnetosphere (AP8MIN). Indeed, thanks to the orbit of the SVOM satellite and the shielding of MXT camera, the overall contribution of solar particles, cosmic rays and trapped electrons is negligible with respect to protons, which are trapped in the South Atlantic Anomaly. This simulation does not take into account damage induced by secondary particles produced inside the spacecraft and whose fraction of energy is deposited in the detector. 

For a nominal mission duration of 3 years, our prediction for the end of life 50~MeV protons equivalent dose corresponds to 6$\times$10$^{9}$ protons.cm$^{-2}$.

\subsection{Expected radiation induced effects on the pnCCD}

Following proton irradiation, the energy resolution of a pnCCD is expected to be degraded due to both dark current increase and charge transfer losses. Firstly, proton irradiation creates defects acting as impurities in the semi-conductor band gap, causing the generation of additional electron-hole pairs. Consequently, a dark current increase is expected and has been measured in pnCCDs after proton irradiation. For illustration, \cite{Meidinger_XMM} report a steep increase of dark current by a factor ~300 after 10~MeV protons irradiation with fluence of $5.8 \times 10^{9}$ protons.cm$^{-2}$ for a 3 $cm^2$ large pnCCD made of the same material, in the same fabrication site that MXT pnCCD, and measured at ambient conditions. We expect such a dark current increase in the MXT pnCCD with visible impact on the dark noise. Additionally, radiation causes displacement damage, which results in an increase of the trapping sites in the silicon lattice, and consequently an increase of the probability of trapping \citep{Meidinger_XMM, Struder, CCD_non_temp_dep}. 
It is known \citep{Meidinger_XMM, Meidinger_trap_types, Meidinger_irrad_conditions} that the performances of pnCCDs are affected principally by two kinds of defects creating energy states in the band gap of silicon, i.e. A-centers and divacancies, which are complexes composed of a vacancy and an oxygen atom, and two vacancies respectively. They are created at different rates according to the concentration of oxygen atoms in the device (A-centers) and to the irradiation dose (divacancies). The defects density determines the amplitude of the transfer inefficiency, while the defect type influences the temperature dependency of the charge loss; they also define all the capture and emission rates. Furthermore, on top of the temperature and defect types, the CTI also greatly depends on the number of signal charges in the pixel volume (i.e. the incoming photons energy).

The charges created by photoelectric absorption of the incident photon in the sensor are likely to be trapped during their transfer to the anodes or, most probably, during their storage for the readout phase. The Charge Transfer Inefficiency (CTI) quantifies this relative statistical charge loss between the readout of two rows. The reconstructed energies are thus shifted towards lower values, and this shift increases with the number of transfers during the image read-out. On top of this effect, the CTI also causes a distortion of the spectral lines, more specifically an enlargement, and a disruption of their symmetry. As a statistical effect, the CTI can be partially corrected, but in the end it adds some uncertainty error to the signal \citep{CCD}. Though measured at very small values on non-irradiated pnCCDs (CTI $\sim$ 10$^{-5}$ on the non-irradiated MXT camera, see \cite{Benjamin}), it has been shown to significantly increase after irradiation, up to values CTI $>$ 10$^{-4}$, as reported by \cite{CCD_irrad_temp_and_flux} on a different type of silicon-based CCD, with measurements of CTI of 2$\times$10$^{-4}$ after irradiation with 40 MeV protons and a 6.4$\times$10$^9$ protons.cm$^{-2}$ fluence. Regarding spectral resolution, \cite{Meidinger_eROSITA} observed on eROSITA pnCCD, whose design, pixel geometries and timings for readout are very similar to MXT, an enlargement of the spectral line at 5.9~keV of 60\% after irradiation with 10 MeV protons and a 5.6$\times$10$^8$ protons.cm$^{-2}$ fluence.
 
Lastly, when pixels are damaged by irradiation, they can produce locally enhanced dark current, significantly higher than their neighboring pixels. Such pixels are called "hot pixels". An increase in hot pixels number has been extensively observed after irradiation in CCDs, for instance on the Hubble Space Telescope \citep{hot_px_HST}. Such events were not reported with pnCCD technology in previous studies but shall be verified on the MXT camera in the operating temperature conditions in Section \ref{sect:results}.

\subsection{Proton irradiation campaign}

\begin{figure}[h]
\begin{center}
\begin{tabular}{c}
\includegraphics[height=7.5cm]{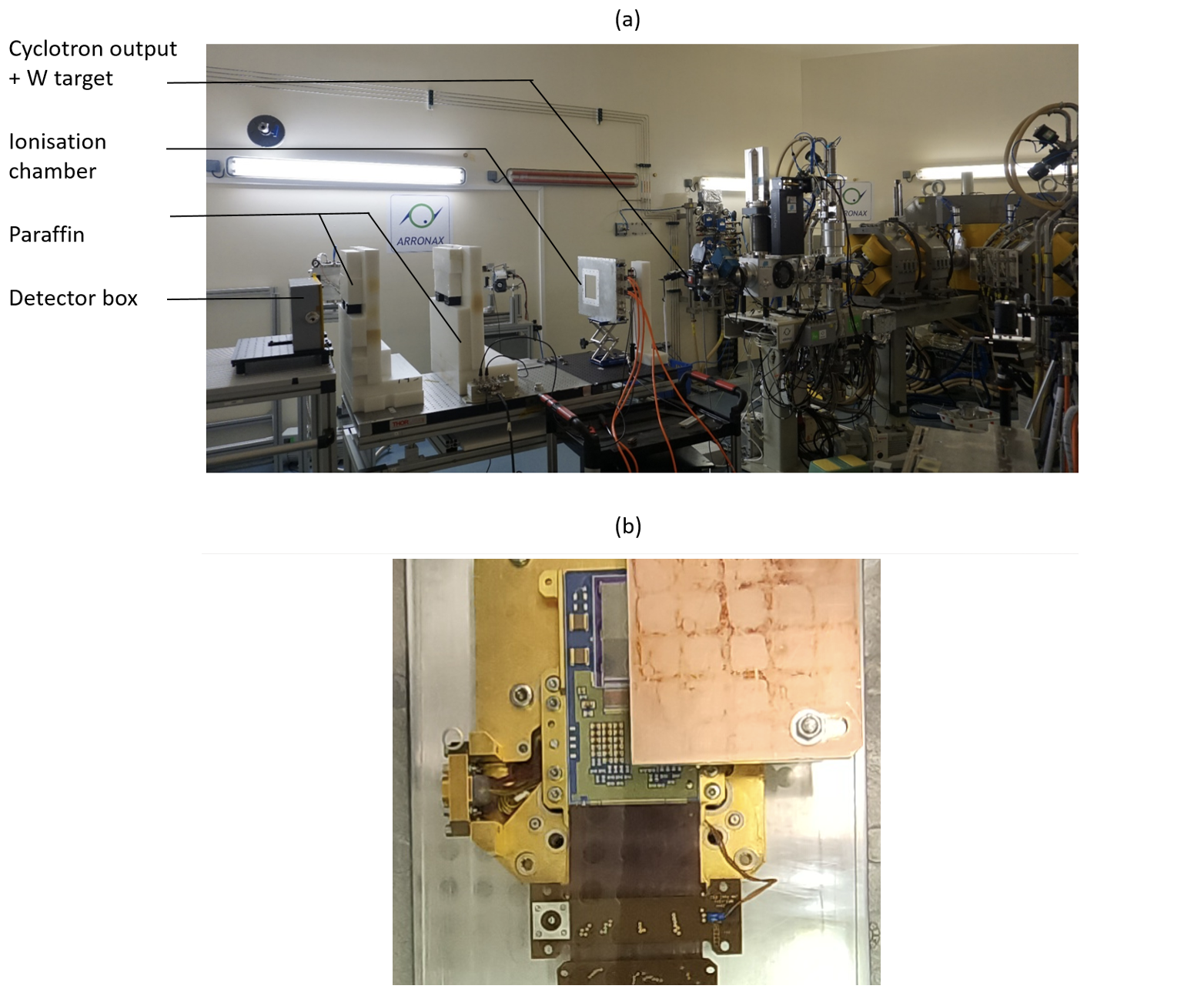}
\end{tabular}
\end{center}
\caption 
{ \label{fig:Arronax} (a) Set-up of the experiment room in Arronax cyclotron to irradiate the detector with 50 MeV protons. The real time proton flux is recorded thanks to a ionisation chamber in the beam path in order to precisely evaluate the fluence. The blocks of paraffin shape the beam to the area to irradiate and also aim at stopping secondary particles like alpha particles and neutrons. (b) Detector in the detector box, with a 6 mm-copper plate mounted just above the sensor guaranteeing that only one half of the detector is irradiated (image area, frame store and associated Camex ASIC). } 
\end{figure}

A proton irradiation campaign was set up at the Arronax cyclotron facility in Nantes (France) in June 2022. A spare model of MXT focal plane with a detector from the same batch as the flight detector was placed in a testing box. The bottom part of the proton shield was removed and replaced by a 6 mm copper slab on half the size of the focal plane in order to directly irradiate only half of the surface of the detector as well as the CAMEX ASIC reading this half part. The goal is, on one hand, to create uniform defects in the illuminated silicon bulk to achieve a single value of equivalent displacement damage dose and, on the other hand, to fully stop the protons on one area of the detector (and CAMEX) to keep a reference during the performance characterisation. The testing configuration is shown in Figure \ref{fig:Arronax}. The beamline delivers protons that are scattered in the experiment room with a tungsten target to produce a large and uniform beam, which is then collimated by blocks of paraffin to produce the desired shape, i.e. a rectangle shaped beam spot of 4 cm x 4 cm with $>$ 95~\% flux uniformity. The proton energy is degraded after the passage in air, in the ionisation chamber used for real-time flux monitoring and in the plexiglas protection cap of the detector box. The beamline is hence configured in order to provide 50 MeV protons at the entrance of the detector (from 68 MeV protons at the output of the cyclotron). The detector is not operated during irradiation: this is representative of the life cycle of MXT camera since radiation damage mainly occurs in the South-Atlantic Anomaly when the focal plane is switched off (contrarily to all other systems kept on to ensure thermal stability).

For simplicity, the irradiation took place at room temperature. Although it may happen that the type of traps generated in CCD detectors depends on the irradiation temperature \citep{p_irrad_temp, CCD_irrad_temp_and_flux}, the most common trap types for pnCCD do not, as suggested by the study made by \cite{Meidinger_XMM} (see also \cite{CCD_non_temp_dep}, \cite{CCD_irrad_temp_and_flux} for similar conclusions on CCDs). 
The flux of 2$\times$10$^{5}$~protons.cm$^{-2}.s^{-1}$ was defined as a compromise between the duration of the tests (8 hours) and the accelerated ageing that could be reported with other semiconductor detectors with high proton flux. Much higher fluxes are generally applied with CCDs \citep{CCD_irrad_temp_and_flux, CCD_higher_flux}.

\subsection{Post-irradiation characterisation}

To cover the entire MXT energy range, two types of X-ray sources have been used. We describe the experimental set-up for each of these characterisation campaigns hereafter. The X-ray fluence was fixed for each of our characterisation campaigns. For the effect of fluence on the CTI and other detector parameters, the reader should refer to \cite{Meidinger_irrad_conditions, CCD_higher_flux, CCD_irrad_temp_and_flux}

\subsubsection{X-ray tube characterization}
\begin{figure}[h!]
    \centering
    \includegraphics[height=6.2cm]{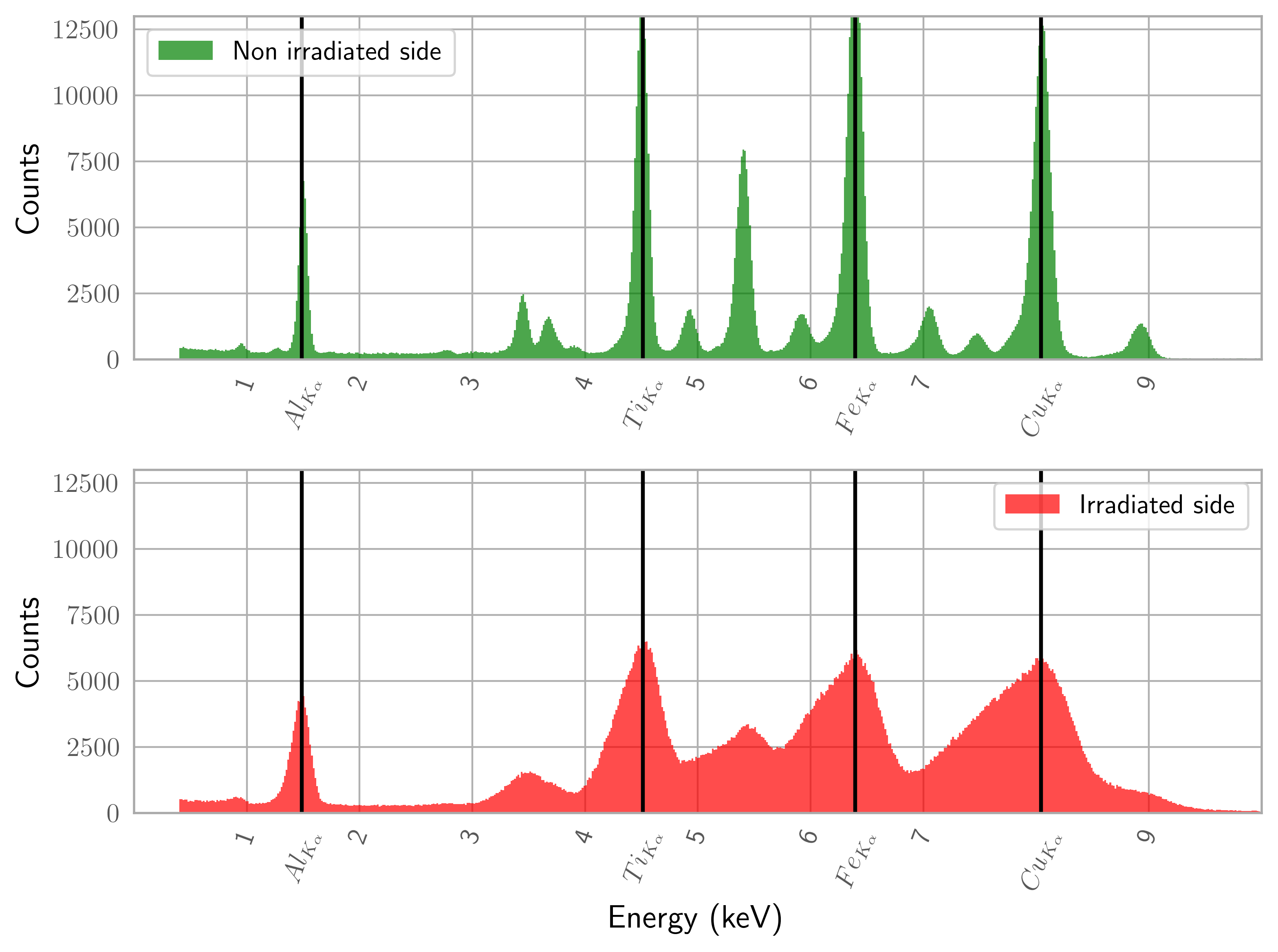}
    \caption 
    { \label{fig:degradation}
    Spectra of the non-irradiated side (top panel, in green) and the proton irradiated side (bottom panel, in red) of MXT pnCCD detector. The spectra were obtained simultaneously with the \textit{Xfluo} source, whose main spectral lines are indicated by black vertical lines. } 
\end{figure}

After irradiation, the focal plane was placed in its test cryostat to be operated between $-90^\circ$C and $-60^\circ$C. A sensor placed on the cold interface of the detector board allows for precise measurements of the detector operating temperature, and ensures a thermal control within $\pm$0.5°C during the full measurement. This configuration allowed us to characterize the dark noise, originating from the leakage current in the semiconductor detector as a function of the temperature. The laboratory setup is also equipped with a full-custom X-ray source called \textit{Xfluo}, composed of an X-ray tube and X-ray fluorescence targets to provide multiple X-ray lines on the full energy range of MXT. A first operation session of the detector in July 2022 at the nominal in-flight temperature of $-65^\circ$C with this source allowed us to verify the functionality of the pnCCD and to observe a clear degradation of the spectral resolution between the non-irradiated and the irradiated parts (see Figure \ref{fig:degradation}). Then, the \textit{Xfluo} X-ray source has been employed to characterize MXT camera response especially in the higher energy band. CTI values were calculated for three main characteristic X-ray lines provided by the source: $Al$ $K\alpha$ at 1.5 keV, $Ti$ $K\alpha$ at 4.5 keV, and $Cu$ $K\alpha$ at 8.0 keV. When in place, the \textit{Xfluo} source illuminates uniformly the whole detector surface, with a flux of about 9000 photons.s$^{-1}$.

\subsubsection{\textit{SOLEIL} characterization}
The lowest energy well exploitable spectral line produced by the \textit{Xfluo} source is the $Al$ $K\alpha$ line at $\sim$1.5~keV. To explore the spectral response of the pnCCD down to 0.2 keV, we applied for beam time to the X-ray metrology beamline of the \textit{SOLEIL} synchrotron facility. The goals of this campaign were to: (i) characterize the spectra on the irradiated and non-irradiated parts of the detector, (ii) measure the gain and CTI on the instrument most critical effective energy range (below 2~keV) and with a monoenergetic source, (iii) measure the detector spectral resolution, and (iv) determine the low energy threshold of the detector. We took advantage of a preparatory test campaign performed in November 2021 for the technical configuration of our cryostat and the beamline, as detailed in \cite{Aline_SOLEIL}. Indeed, the SOLEIL beam settings are very peculiar for this experiment since we need a photon flux 6 to 7 orders of magnitude lower than the nominal operation of the beamline, in order to limit the photon pile-up in our images of 100 ms integration. 

\begin{figure*}[h]
\begin{center}
\includegraphics[height=9.5cm]{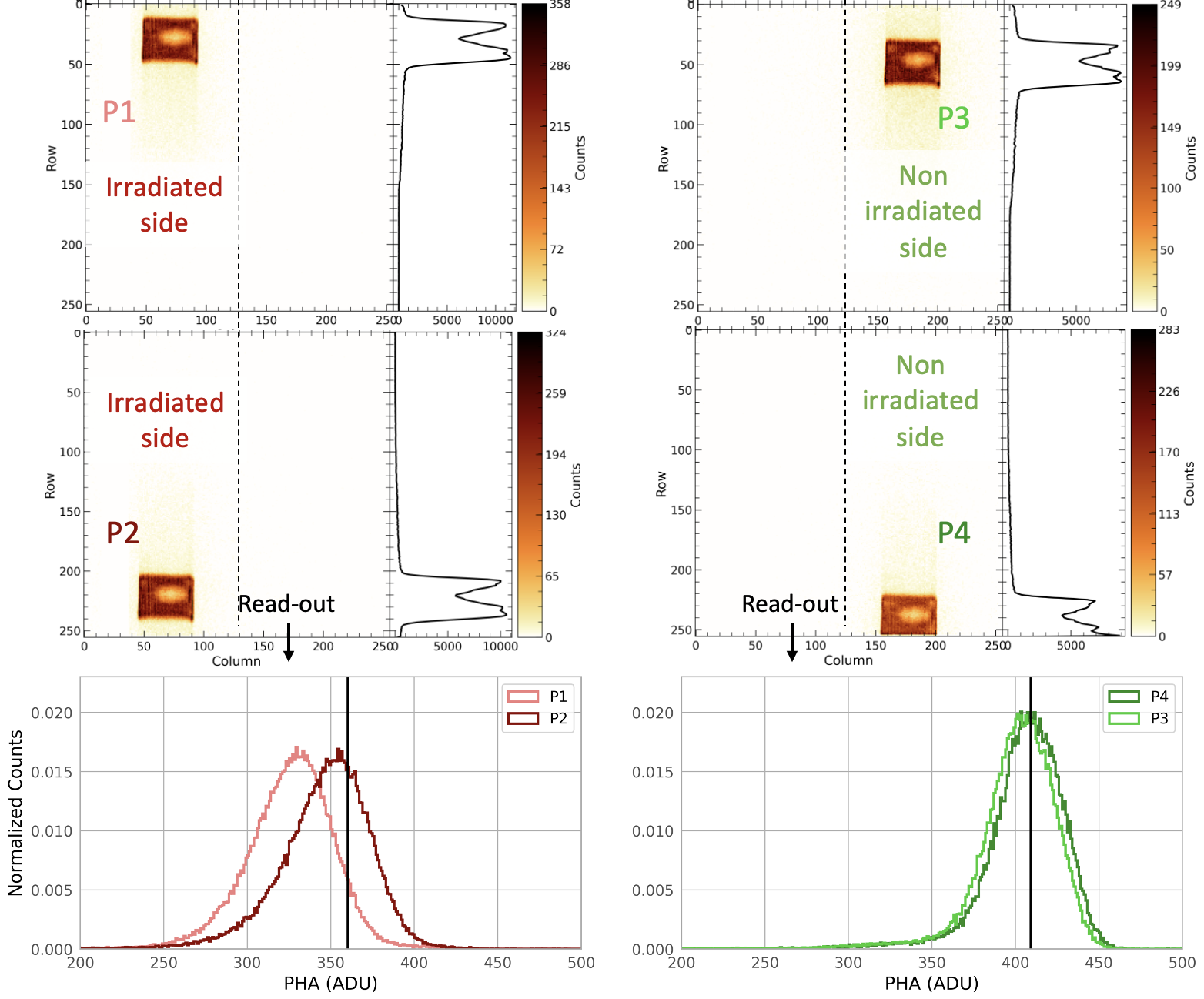}
\end{center}
\caption 
{ \label{fig:4_spectra}
Results of data acquisition at \textit{SOLEIL} at 900~eV. The count rate is lower in the center of the beam patch due to pile-up. Here, the spectra from four different beam positions on the detector matrix are shown. On the irradiated side (left), the effect of the CTI is clearly visible: the spectral lines are wider compared to the non-irradiated side, and shifted to lower energies (compared to the input energy, shown as a black vertical line). The effect is more important for the top of the detector matrix (P1 position) with respect to the bottom part (P2), where only the frame store contributes to the CTI. The output node is at row number 255. } 
\end{figure*}

The beam line was successfully tuned with an excellent spectral purity, and the produced energies input ranged from 200~eV to 1900~eV. The beam was tuned to its largest divergence, creating a patch on the detector matrix -- a $\sim$~3$\times$2.2~mm$^2$ rectangle covering roughly 40 pixels along the columns and 30 pixels along the lines, as can be observed in Figure \ref{fig:4_spectra}. The flux on the beam patch is about 150 photons.s$^{-1}$. Inside the cryostat, at -65$^\circ$C (on the CCD), the focal plane assembly is mounted on two micrometric displacement tables. This allows the displacement of the detector with respect to the beam spot while maintaining the mechanical alignment of the cryostat with the beamline chamber to operate under vacuum. The displacement is useful, on one hand, to illuminate alternatively the non-irradiated and the irradiated part of the detector and, on the other hand, to illuminate the irradiated part from the top to the bottom to characterize the charge transfer inefficiency (see Figure \ref{fig:4_spectra}). We did not have sufficient beam time to fully characterize the non-irradiated part during this campaign. A measurement at each energy, as well as a thorough pavement with the beam patch along the detector rows with an 900~eV input energy to measure the CTI, were conducted for consistency.

\subsection{Data Processing}
\label{sect:data_processing}

\subsubsection{Determination of spectral calibration parameters: offset, gain, CTI}
During all the characterisation campaigns, the MXT detection chain is operated in {\it event} mode, which is the nominal in flight mode. The pipeline of X-ray event extraction consists of four steps: pattern recognition, energy calibration, charge sharing correction, and photon list building. The details for each of these steps is described in \cite{Benjamin}.

However, in particular for the proton irradiated data, the energy at this final step is not fully correct, as it is not corrected for CTI. The CTI is a row-wise effect. Its measurement is derived from the photon energy measured against the number of transfers - that is, the number of rows the charge has to pass to reach the read-out electronics. In order to increase the statistical accuracy, the measurement of the CTI is not derived row by row, but based on the sum of spectra along groups of rows. The number of rows along which the spectra was summed to measure the CTI has been optimized - the more rows we group for a single energy measurement of a line, the more precise this energy measurement will be, but the less points we will have to sample the charge loss effect along the whole length of the matrix. The measurement of the CTI can be operated on a single \textit{Xfluo} dataset, as the \textit{Xfluo} illuminates the full detector matrix. For data obtained at \textit{SOLEIL}, however, the beam patch only covers a portion of the rows. In order to precisely quantify the CTI, our data taking strategy here was thus to cover a majority of the rows with the beam patch, as is illustrated Figure \ref{fig:4_spectra}, to then concatenate those datasets and finally apply a data processing similar to that of \textit{Xfluo}.

\begin{figure*}[]
    \centering
    \includegraphics[scale = 0.4]{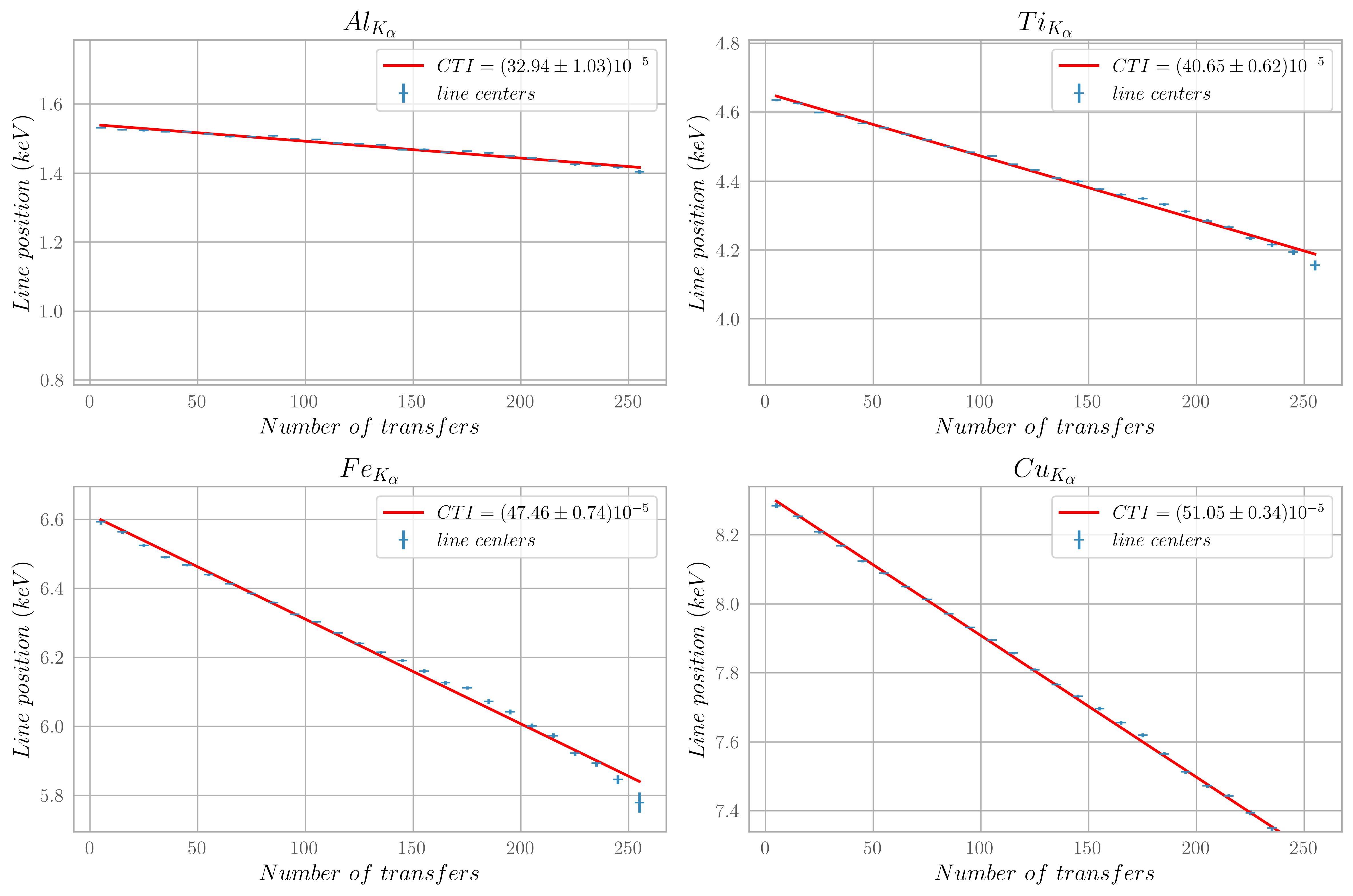}
    \caption{Examples of linear regressions of the spectral centroid position for the main lines in \textit{Xfluo} raw calibrated data against the number of transfers to determine the (image area) CTI. The rows are grouped by 10. } 
    \label{fig:CTI}
\end{figure*}

The energy line centroids are fitted as a function of the corresponding rows of the detector matrix. Assuming the charge loss along the rows is linear, we then perform a linear regression. Examples of such datasets are shown Figure \ref{fig:CTI}. We measure the slope of the linear regression, called $\alpha$ (in eV per transfer), and divide it by the theoretical energy value of the line, noted $E_0$ (in eV). This gives the ratio of charge loss per transfer over total charge before any transfer, and is how we determine our CTI values, as expressed by equation \ref{eq:CTI}. The error on the CTI value is measured as the standard error of the estimated slope, under the assumption of residual normality.

\begin{equation}
\label{eq:CTI}
    CTI = \frac{\alpha}{E_0}
\end{equation}

Each event requires between 256 and 512 transfers to be read, but we have access to the row-wise energy shift in the image area only. Consequently, we use this formula to estimate the CTI in the image area and we estimate the energy loss in the frame store by energy calibration after CTI correction. A gain loss in ADU/eV is expected after irradiation. In this paper, all CTI values presented correspond to CTI in the image area.

\subsubsection{Iterative approach to calibration}
In the context of spectral calibration of the MXT camera, the CTI correction must be implemented in the data analysis pipeline. The CTI correction is applied column by column, following the equation: 
\begin{equation}
\label{eq:CTI_correction}
    E_{corrected}(j) = \frac{E_{raw}(j)}{CTE^{255-j}}
\end{equation}
where the corrected energy $E_{corrected}$ at row $j$ corresponds to the original raw energy $E_{raw}$ at the same row, rectified by the Charge Transfer Efficiency ($CTE = 1 - CTI$) weighted by the corresponding number of transfers ($255 - j$). Gain and offset are determined per column, while the CTI determination is row-wise, making these two corrections interdependent. Therefore, an iterative approach is required to achieve a fully calibrated set of parameters (see chapter 4.3.2 of \cite{Benjamin_thesis} PhD thesis for more detail). The process starts with an initial iteration (iter0) where a first set of calibration parameters is derived, assuming the CTI is zero (CTI$_0$ = 0). Using the energy-calibrated data, a preliminary CTI value (CTI$_1$) is then calculated. In the next iteration (iter1), this newly determined CTI$_1$ is applied to correct the raw ADU data, and a second set of calibration parameters is derived from the CTI-corrected data. This iterative process continues until the CTI value becomes sufficiently small — specifically, when it falls below $10^{-5}$. 

For a non-irradiated detector, the iterative process typically converges after the first iteration. However, for an irradiated detector, up to six iterations may be required to reach the desired accuracy. Since the CTI is energy dependent, the CTI must be derived for several energies across MXT energy range and a function CTI(E) is defined in each iteration and applied to the data as described above.

\section{Experimental results and spectral performance}
\label{sect:results}

\subsection{Effects of proton irradiation on the dark noise}
\begin{figure}[!h]
    \centering
    \includegraphics[scale = 0.37]{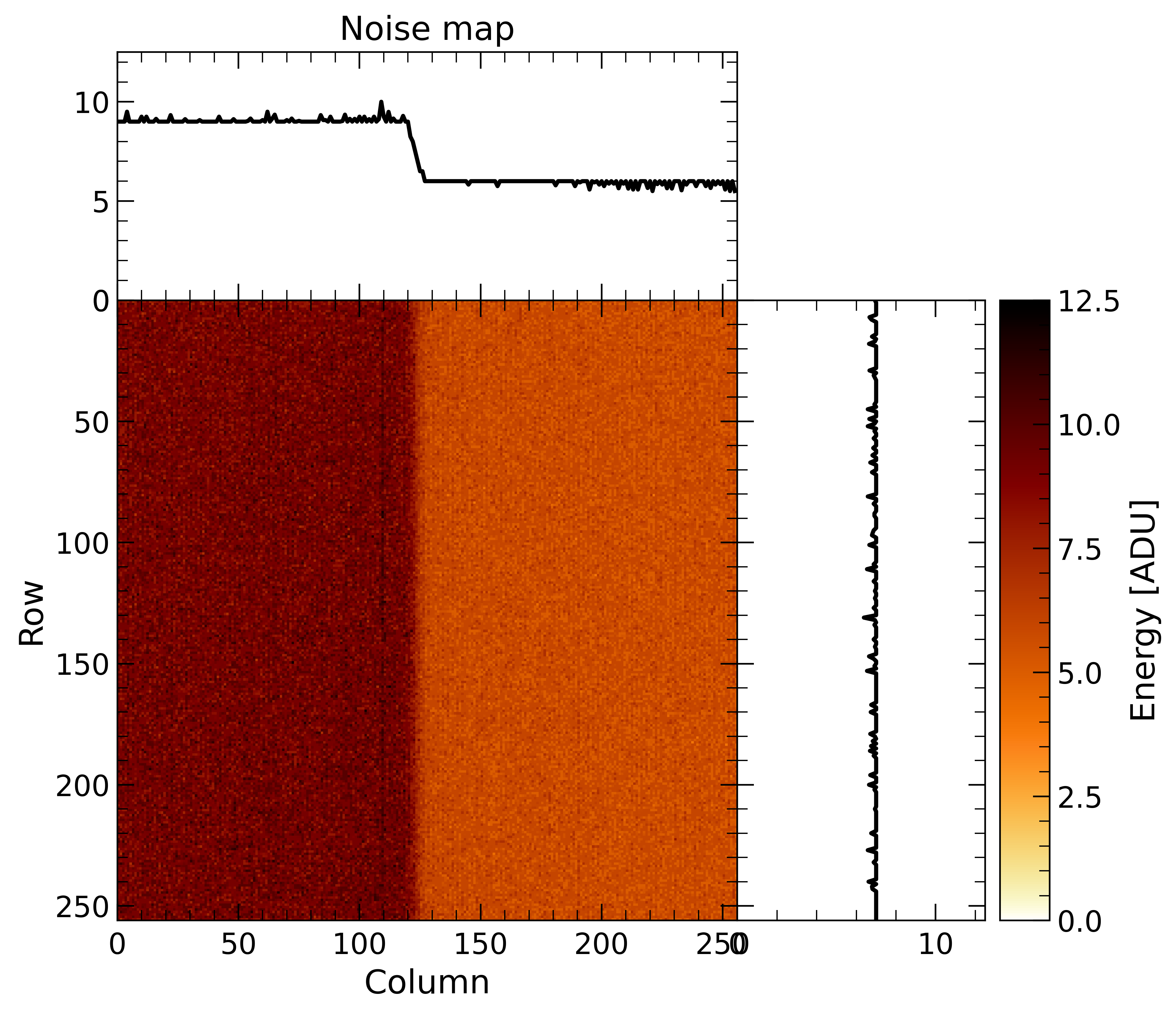}
    \caption{Noise map taken of the irradiated detector. The increase of the dark noise due to irradiation is clearly visible on the left side of the detector matrix.}
    \label{fig:dark_noise}
\end{figure}

\begin{figure}[!h]
    \centering
    \includegraphics[scale = 0.48]{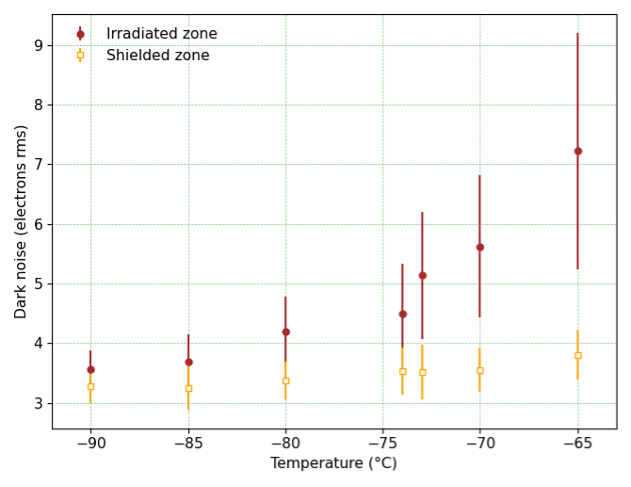}
    \caption{Dark noise measurements as a function of the temperature with average value for each of the detector zones. The error bars are the standard dispersion over all the pixels. The values are converted in eV after spectral calibration done from \textit{Xfluo} data.}
    \label{fig:dark_noise_increase}
\end{figure}

Proton irradiation causes a uniform increase in dark noise (see Figure \ref{fig:dark_noise}). Its level is extracted from the so called {\it dark measurement} procedure, in \textit{full frame} mode and without any photon illumination, to get a measurement of the noise level of every pixel of the detector matrix. This noise value is shown as a function of temperature in Figure \ref{fig:dark_noise_increase}.

Figure \ref{fig:dark_noise_increase} suggests that the dark noise increases significantly with temperature, which leads to the interpretation of an increase of dark current in the detector caused by thermal generation of electron–hole pairs. This is because protons generate deep level defects, which can behave not only as traps for the electrons, but also as generation centres for electron-hole pairs \citep{Meidinger_XMM}, and this directly translates in an increase in the dark noise. We do not expect an additional noise excess due to the ASIC itself because the total ionising dose that could degrade analog parameters of the ASIC is really low in the experiment and during the mission ($ < $ 10 krad).

Additionally, after irradiation, when applying the nominal operating voltages, we failed in properly bias the JFETs at the anodes of each column of the sensor, which are responsible for the first stage of amplification of the detection chain. We had to change the potential of the guard ring anode and the potential of the reset anode to evacuate the excessive charges accumulated by thermal generation around the JFET structures. The Guard ring anode was moved from -18V to -11V, knowing that the phase voltages are between -16 and -22V. Reset anode was changed from 0V to +30V. We noticed this behavior as a threshold effect: as long as the leakage current is below some acceptable value, the nominal bias voltages are valid. Above this value, two bias voltages have to be strongly modified to properly switch on the detector. This result is extremely important for the operation in flight of the detector. The increase in leakage current will be monitored by regular in-flight dark measurements.

\begin{figure}[!h]
    \centering
    \includegraphics[scale = 0.34]{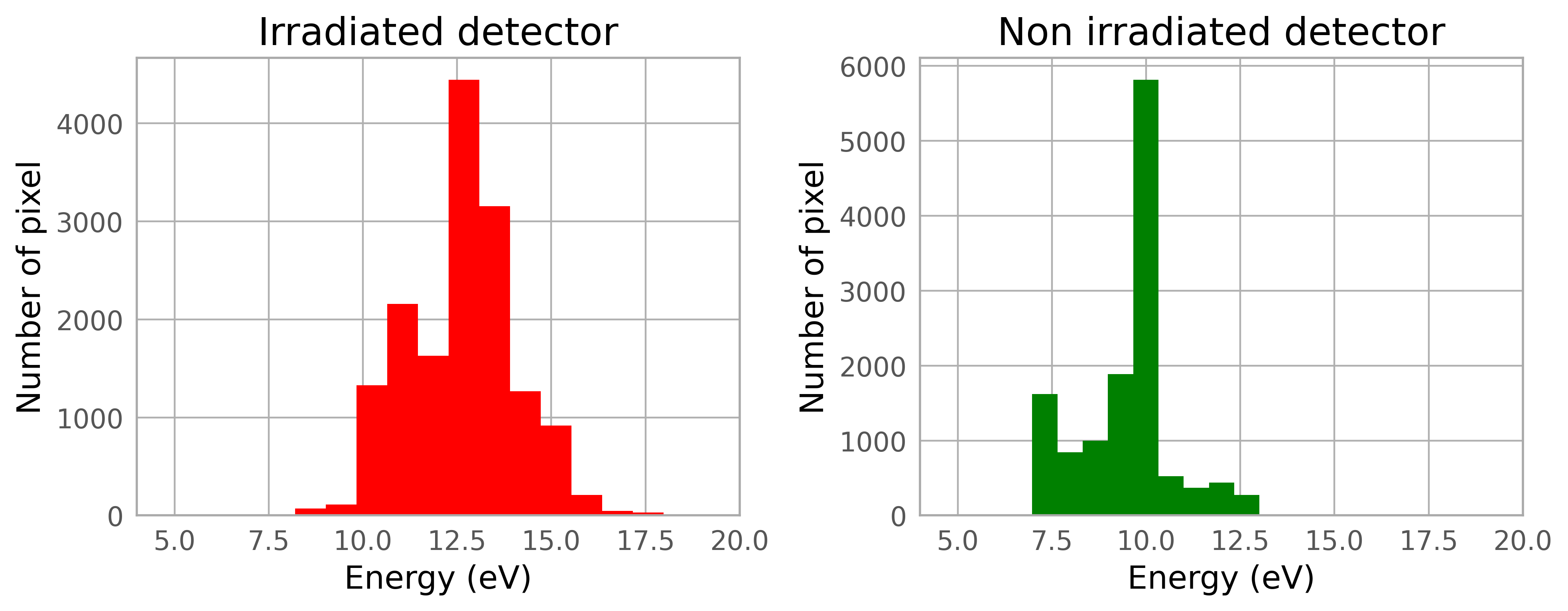}
    \caption{Noise histograms taken from \textit{SOLEIL} {\it dark measurements}. }
    \label{fig:hot_px} 
\end{figure}

Lastly, we have been looking for the appearance of hot pixels after irradiation, and none was found. Hot pixels would appear as outliers in the pixels noise energy distribution. The histograms of \textit{SOLEIL} noise maps, Figure \ref{fig:hot_px}, display no such outliers of the dark image signals for any pixel.

\subsection{Charge Transfer Inefficiency}
Using both the \textit{SOLEIL} and \textit{Xfluo} sources illuminating MXT irradiated detector, we measured the CTI according to a method adapted to each campaign, as presented in Section \ref{sect:methods}. In both cases, we observe, as expected, an increase of CTI due to proton irradiation, by an order of magnitude, consistently with the study on other detectors presented in the last Section. 

The CTI on the non-irradiated part of the detector at \textit{SOLEIL} (($5.22 \pm 0.26)~10^{-5}$ at 900~eV) is consistent with the CTI of the non-irradiated flight model (($4.86 \pm 0.09)~10^{-5}$ at the same energy), that has been previously characterized during MXT Panter calibration campaign \citep{Benjamin}, though slightly higher. Furthermore, it is also consistent with the one measured on the same detector before irradiation in similar conditions (($5.2 \pm 0.7)~10^{-5}$ at 1000~eV), during the previous \textit{SOLEIL} campaign, detailed in \cite{Aline_SOLEIL}. As a result, we assume that the new spare model detector response is fully due to our proton irradiation. 

The CTI measurements obtained from \textit{SOLEIL} data are reported Table \ref{tab:CTI}.

\begin{table}[h]
\caption{CTI results on the irradiated detector from {\it SOLEIL} data (at -65$^\circ $). The CTI measurement we obtained during this campaign for the non-irradiated side is ($5.22 \pm 0.26)~10^{-5}$ at 900 ~eV}
\label{tab:CTI}
\begin{center} 
\begin{tabular}{c|c}
\rule[-1ex]{0pt}{3.5ex} $Energy$ (eV) & $CTI (10^{-5})$ \\
\hline
\rule[-1ex]{0pt}{3.5ex} $200$ & $27 \pm 7$  \\
\hline
\rule[-1ex]{0pt}{3.5ex} $300$ & $35 \pm 4$\\ 
\hline
\rule[-1ex]{0pt}{3.5ex} $500$ & $37 \pm 2$\\
\hline
\rule[-1ex]{0pt}{3.5ex} $700$ & $41 \pm 2$\\
\hline
\rule[-1ex]{0pt}{3.5ex} $900$ & $40 \pm 3$\\ 
\hline
\rule[-1ex]{0pt}{3.5ex} $1300$ & $45 \pm 1$\\
\hline
\rule[-1ex]{0pt}{3.5ex} $1550$ & $45 \pm 1$\\
\hline
\rule[-1ex]{0pt}{3.5ex} $1850$ & $48 \pm 1$\\
\end{tabular}
\end{center}
\end{table}

\begin{table*}[]
    \caption{{\it Xfluo} CTI results (in 10$^{-5}$) for the irradiated and non-irradiated detector at different temperatures. The CTI results to compare with the {\it SOLEIL} results are written in bold. } 
    \label{tab:Xfluo_CTI}
    \centering
    \begin{adjustbox}{max width=\textwidth}
    \begin{tabular}{c|ccc|ccc}
        Temp. ($^\circ$C) & \multicolumn{3}{c|}{Irradiated} & \multicolumn{3}{c}{Non-Irradiated} \\
        \hline
        & 1486 eV & 4508 eV & 8040 eV & 1486 eV & 4508 eV & 8040 eV \\
        \hline
        \rule[-1ex]{0pt}{3ex} -65 & \textbf{\bm{$28 \pm 1$}} & \textbf{\bm{$34 \pm 1$}} & \textbf{\bm{$46 \pm 1$}} & $3.8 \pm 0.3$ & $2.5 \pm 0.2$ & $2.5 \pm 0.1$ \\ [0.8ex]
        \hline
        \rule[-1ex]{0pt}{3ex} -75 & $34 \pm 1$ & $41 \pm 1$ & $49 \pm 1$ & $3.4 \pm 0.3$ & $2.4 \pm 0.2$ & $1.8 \pm 0.1$ \\ [0.8ex]
        \hline
        \rule[-1ex]{0pt}{3ex} -80 & $34 \pm 1$ & $40 \pm 1$ & $48 \pm 1$ & $3.0 \pm 0.3$ & $1.3 \pm 0.2$ & $1.4 \pm 0.1$ \\ [0.8ex]
        \hline
        \rule[-1ex]{0pt}{3ex} -85 & $31 \pm 2$ & $32 \pm 1$ & $37 \pm 1$ & $2.2 \pm 0.4$ & $1.5 \pm 0.3$ & $1.2 \pm 0.1$ \\ [0.8ex]
        \hline
        \rule[-1ex]{0pt}{3ex} -90 & $26 \pm 1$ & $26 \pm 1$ & $31 \pm 1$ & $1.1 \pm 0.4$ & $1.6 \pm 0.2$ & $1.1 \pm 0.1$ \\ [0.8ex]
    \end{tabular}
    \end{adjustbox}
\end{table*}

The results for both the irradiated and non-irradiated sides of the detector obtained in the laboratory with the X-ray source are summarized in Table \ref{tab:Xfluo_CTI}. We note that the CTI measurements with the laboratory X-ray source are significantly different from the ones obtained with \textit{SOLEIL} data. For instance, at $\approx$ 1.5 keV, we measure a CTI of $(45 \pm 1)~10^{-5}$ from \textit{SOLEIL} data, whereas we measure a CTI of $(28 \pm 1)~10^{-5}$ (at -65$^\circ$C) with the laboratory with the X-ray source. We discuss the possible origins of this discrepancy in Section \ref{sect:discussions}.

\subsection{CTI modelling and correction}

\begin{figure}[h!]
    \centering
    \includegraphics[scale = 0.42]{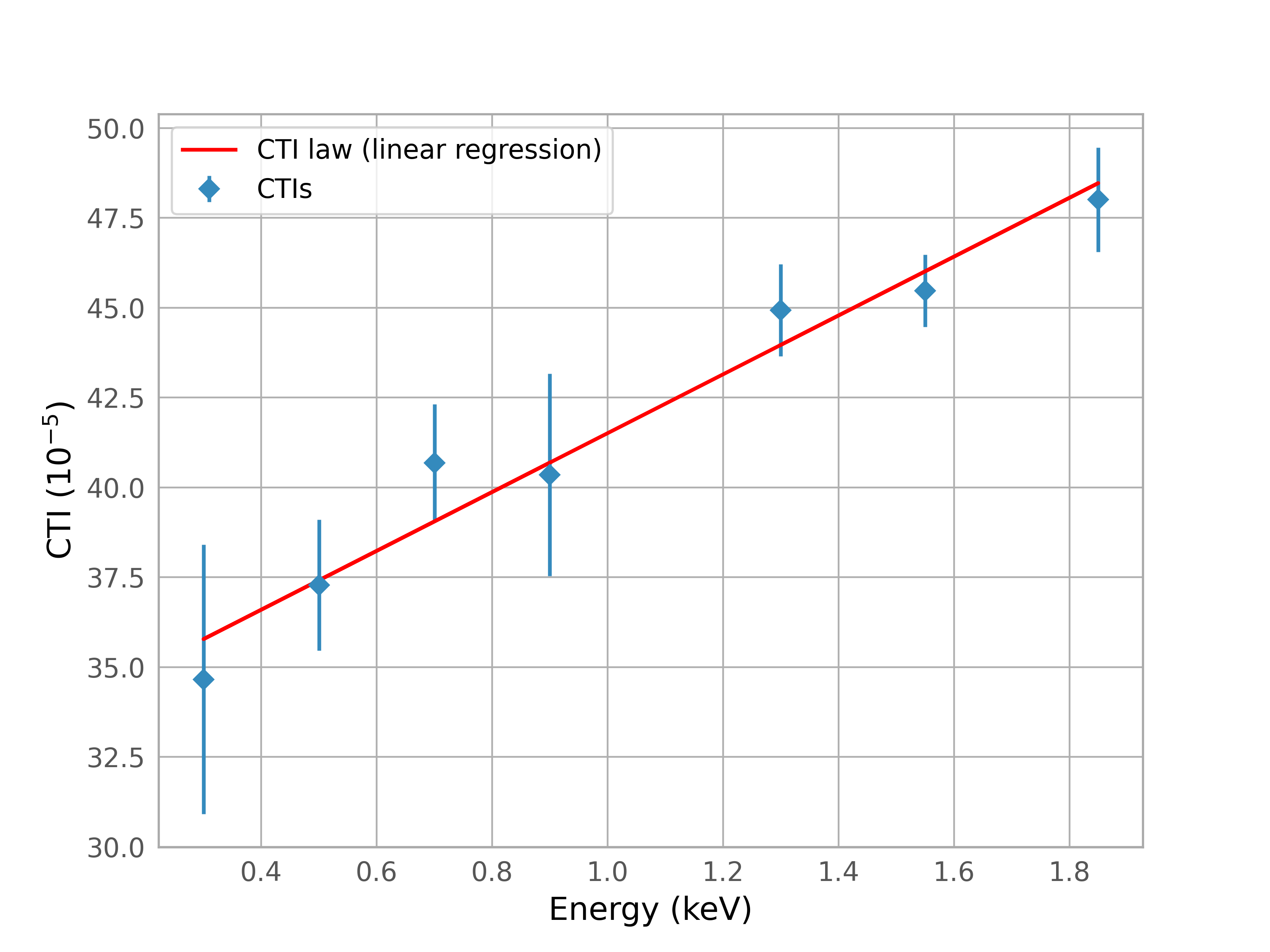}
    \caption{CTI results from \textit{SOLEIL} data (-65$^\circ$C) of the irradiated part of the detector as a function of energy. The linear regression is the chosen CTI correction law for the data processing. }
    \label{fig:CTIvsE}
\end{figure}

In addition to the CTI increase of one order of magnitude (i.e. from $(5.2 \pm 0.3)~10^{-5}$ to ($40 \pm 3)~10^{-5}$ at 900 ~eV according to \textit{SOLEIL} data), after irradiation, we observe an inversion in the trend of the CTI with respect to the energy. Indeed, as is shown in Figure \ref{fig:CTIvsE} and Figure \ref{fig:CTIvsE_genX} for \textit{SOLEIL} and \textit{Xfluo} respectively, the CTI after proton irradiation increases with energy, whereas it was observed to decrease with energy before irradiation \citep{Benjamin, Aline_SOLEIL}.

For both campaigns, our CTI measurements are well fitted by linear laws. As such, a linear law was applied to correct data for CTI effects across the full MXT energy band. For \textit{SOLEIL}, Equation \ref{eq:CTI_corr_soleil} is the established CTI correction law, and Equation \ref{eq:CTI_corr_genX} for \textit{Xfluo} (-65$^\circ$C).

\begin{equation}
\begin{split}
    CTI_{\textsubscript{\textit{SOLEIL}}}(E) = (8.19 \pm 0.75 \times E(keV) \\
    + 33.3 \pm 0.85)~10^{-5}
\end{split}
\label{eq:CTI_corr_soleil}
\end{equation}

\begin{equation}
\begin{split}
    CTI_{\textsubscript{\textit{Xfluo}}}(E) = (2.57 \pm 0.24 \times E(keV) \\+ 24.8 \pm 0.44)~10^{-5}
\end{split}
\label{eq:CTI_corr_genX}
\end{equation}

\begin{figure}[h!]
    \centering
    \includegraphics[height=5.8cm]{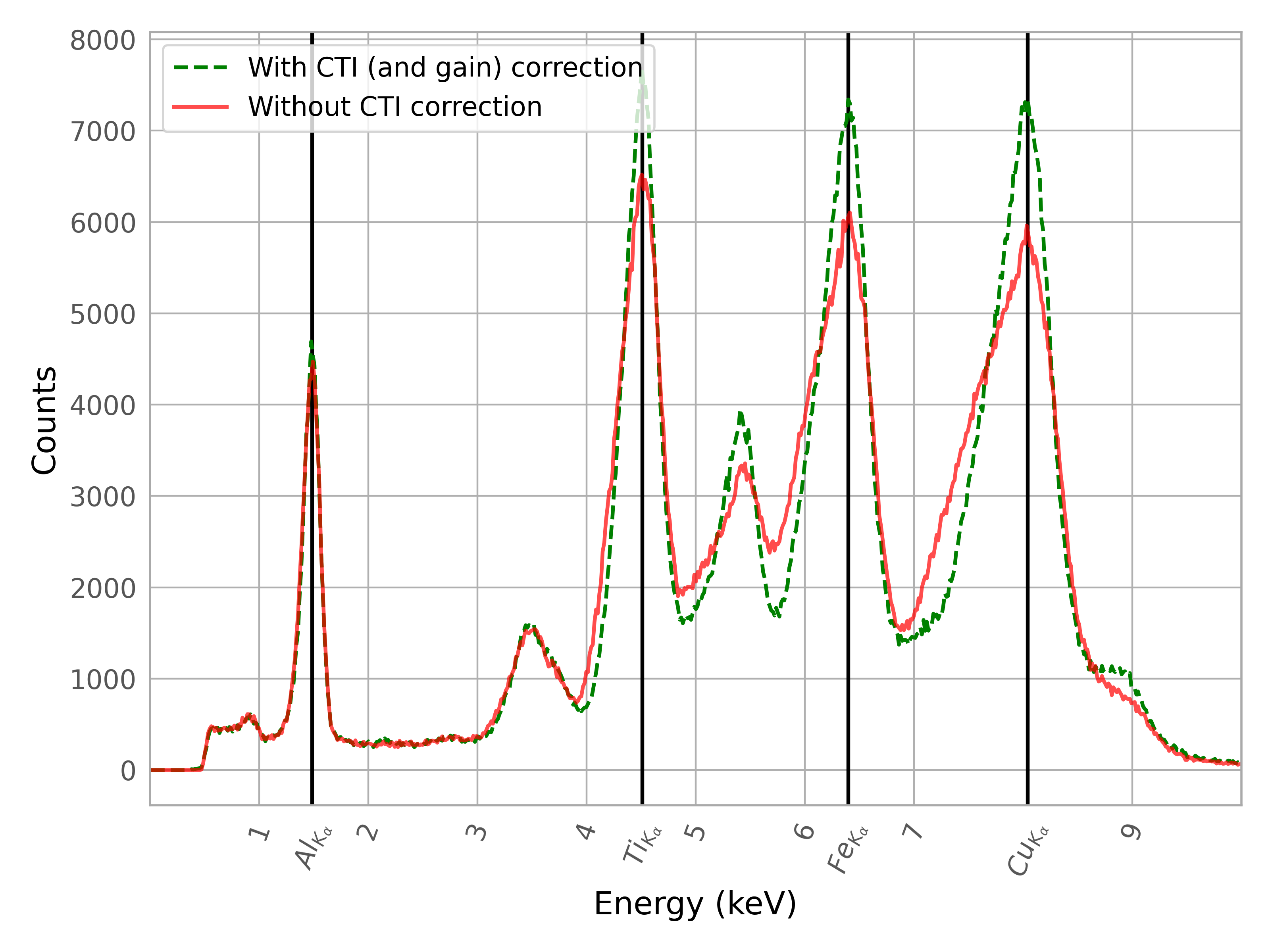}
    \caption 
    { \label{fig:CTI_corr_Xfluo}
    Comparison of spectra from \textit{Xfluo} measurements, before (red filled line) and after (green dashed line) our CTI and gain correction, to highlight the effect of our data processing on the irradiated detector. The energy of the main lines of the \textit{Xfluo} source used for the data processing are represented by black vertical lines. } ¨
\end{figure}

An example of the effect of the CTI correction is shown in Figure \ref{fig:CTI_corr_Xfluo} for \textit{Xfluo} data. From this, the spectral performance of the MXT instrument after irradiation, namely the measurement of the energy resolution, can be established, as described in the following subsections. 

\subsection{Detector gain}
\begin{figure}[!h]
    \centering
    \includegraphics[scale = 0.35]{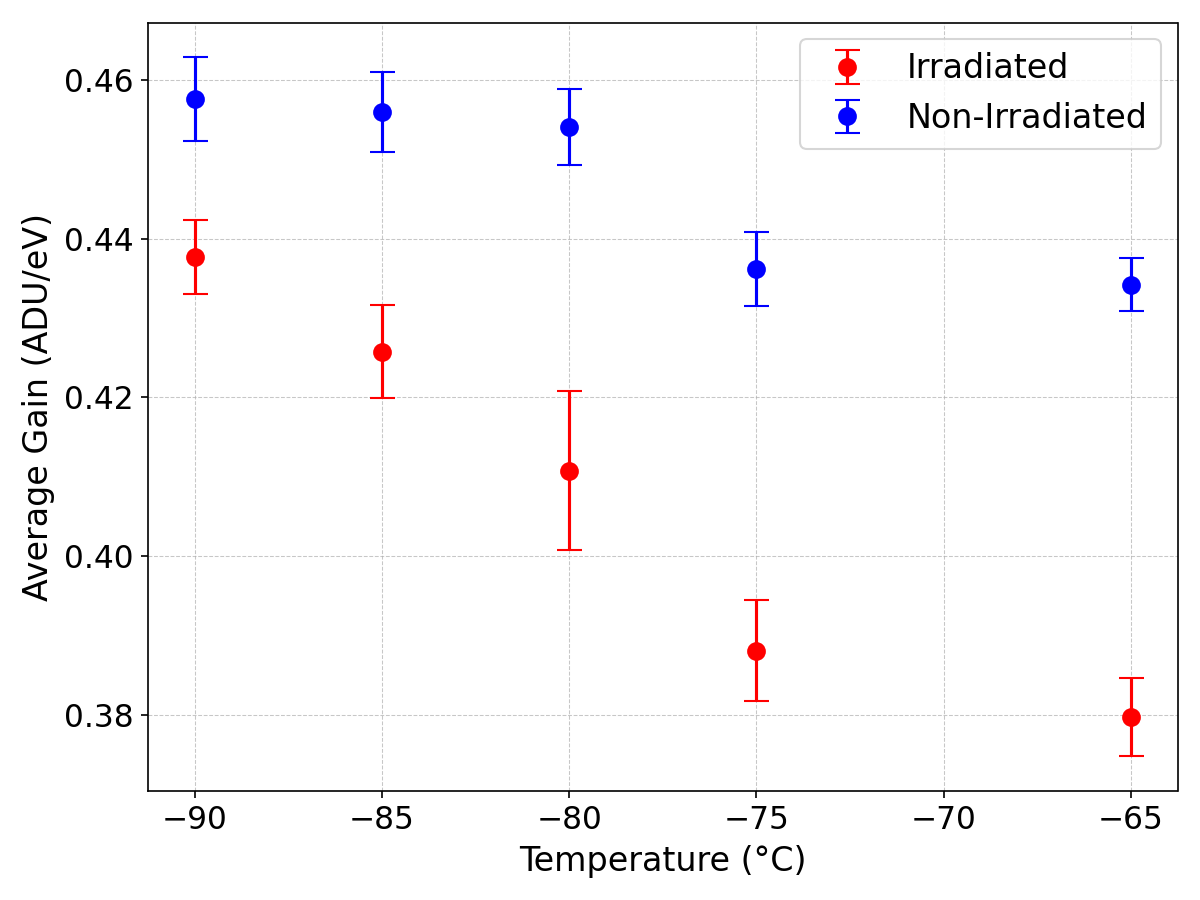}
    \caption{Evolution of the detector gain with temperature for the irradiated and non-irradiated detectors measured with \textit{Xfluo}. Error bars represent the standard deviation along the columns.}
    \label{fig:gainvsT}
\end{figure}

Figure \ref{fig:gainvsT} illustrates the behaviour of detector gain variation with temperature. This gain variation has been measured on CTI corrected data, i.e. corrected from the transfer inefficiency contribution of the image area, showing a clear apparent change of gain. We interpret the gain loss in the irradiated part compared to the non-irradiated part as an increase charge transfer inefficiency in the frame store area. We can notice that the gain loss in the frame store is stronger when the temperature increases. We note that the gain variation between the irradiated and non-irradiated parts is less pronounced at colder temperatures. 

\subsection{Low level threshold}

The LLT averaged over all pixels has been measured before irradiation to be about 46~eV on the flight detector at nominal temperature during the end-to-end calibration campaign at {\it PANTER} facility \citep{Benjamin}. Nevertheless, the dark noise increase observed after irradiation (see Figure \ref{fig:dark_noise_increase}) implies an increase of low-level threshold and possibly loss of detection efficiency for MXT.

\begin{figure}[!h]
    \centering
    \includegraphics[scale = 0.35]{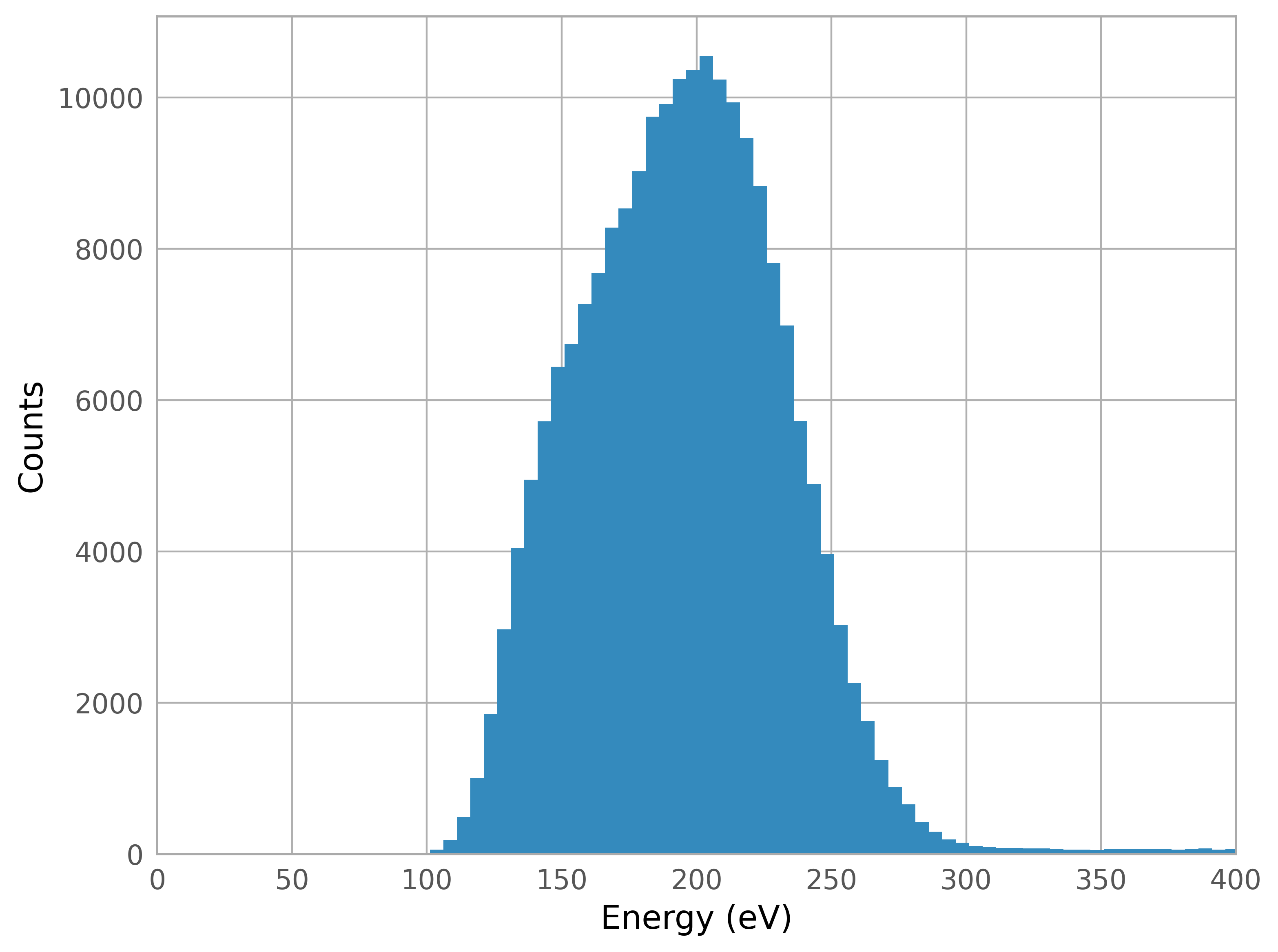}
    \caption{Calibrated and CTI corrected 200~eV spectra obtained from the irradiated detector (\textit{SOLEIL}). The asymmetry of the lower energy tail of the spectral line can be explained by the non-uniform increase of LLT in the irradiated detector, averaging to around 100~eV, and causing a statistical error at low energy. We note that at these energies single events dominate largely over split events, representing 70\% of all events \citep{Benjamin}. }
    \label{fig:spectra_200eV} 
\end{figure}

The average LLT value obtained from the dark noise maps of the irradiated detector is 104~eV at -65$^{\circ}$C (84~eV at -70$^{\circ}$C). This is approximately a factor of 2 larger than the value for the non-irradiated detector. However, this level remains low enough with respect to the 200~eV MXT scientific requirement, as illustrated by the fact that 200 eV spectral line generated at {\it SOLEIL} is still clearly detected, as illustrated in Figure \ref{fig:spectra_200eV}. This indicates that we do not expect any reduction of the nominal MXT spectral band due to the aging of the detector, in accordance with MXT requirements.

\subsection{Energy resolution}

Figure \ref{fig:CTI_corr_Xfluo} highlights the importance of CTI correction before attempting spectral lines fitting, as the raw spectral lines are widened and develop a {\it low-energy tail} due to the radiation damage induced CTI. Due to the asymmetry of the lines, the model we chose to estimate the FWHM is the so-called \textit{Crystal ball} model (see Appendix E of \cite{Crystal_ball}), composed of a Gaussian core and a power law end-tail, well fitting spectral data in presence of energy loss (hence charge trapping). Other fitting methods - including symmetrical and asymmetrical Gaussian fits - were tested on the CTI corrected spectral lines, but yielded less satisfactory fits. An example of a \textit{SOLEIL} spectral line fit can be consulted Appendix A. The energy resolution results of the \textit{SOLEIL} campaign are given Table \ref{tab:FWHM_SOLEIL_cb}.

\begin{table}[ht]
    \begin{center}
     \caption{Energy resolution of the irradiated spare model ({\it SOLEIL} data) according to the \textit{Crystal ball} fitting method. } 
    \begin{tabular}{c|c}
    Energy (eV) & FWHM (eV) \\ 
        \hline
    \rule[-1ex]{0pt}{3.5ex} $200$ & $77 \pm 1$  \\ 
    \hline
    \rule[-1ex]{0pt}{3.5ex} $300$ & $87 \pm 1$\\ 
    \hline
    \rule[-1ex]{0pt}{3.5ex} $500$ & $99 \pm 1$\\ 
    \hline
    \rule[-1ex]{0pt}{3.5ex} $700$ & $109 \pm 1$\\ 
    \hline
    \rule[-1ex]{0pt}{3.5ex} $900$ & $126 \pm 1$\\ 
    \hline
    \rule[-1ex]{0pt}{3.5ex} $1300$ & $140 \pm 1$\\
    \hline
    \rule[-1ex]{0pt}{3.5ex} $1550$ & $151 \pm 1$\\ 
    \hline
    \rule[-1ex]{0pt}{3.5ex} $1850$ & $165 \pm 1$\\
    \end{tabular}
    \label{tab:FWHM_SOLEIL_cb}
    \end{center}
\end{table}

\begin{table}[]
    \centering
    \caption{{\it Xfluo} resolution results, obtained thanks to a Gaussian fit with a quadratic component. }-
    \begin{tabular}{c|c|c}
          & Energy (eV) & FWHM (eV)\\
        \hline
        \rule[-1ex]{0pt}{3.5ex} $Al$ $K\alpha$ & 1,486 & 164 $\pm$ 3\\ [1ex]
        \hline
        \rule[-1ex]{0pt}{3.5ex} $Ti$ $K\alpha$ & 4,509 & 237 $\pm$ 13\\ [1ex]
        \hline
        \rule[-1ex]{0pt}{3.5ex} $Fe$ $K\alpha$ & 6,395 & 306 $\pm$ 14\\ [1ex]
        \hline
        \rule[-1ex]{0pt}{3.5ex} $Cu$ $K\alpha$ & 8,340 & 366 $\pm$ 12\\ [1ex]
    \end{tabular}
    
    \label{tab:Xfluo_G+C_FWHM}
\end{table}

On the contrary, \textit{Xfluo} spectra after CTI correction do not display as asymmetrical lines as for the \textit{SOLEIL} dataset, see Figure \ref{fig:CTI_corr_Xfluo}. This is probably due to the fact that the \textit{Xfluo} source covers higher energies, where the line widths are relatively larger and this effect filters finer distortion effects. Hence, a Gaussian function with an additional quadratic component to reproduce the background flux around the spectral line has been sufficient for the energy resolution determination on \textit{Xfluo} data. An example of a \textit{Xfluo} spectral line fit can be consulted Appendix B. At 1.5~keV, our spectral line fit yields a spectral resolution of 164$\pm$3~eV; our energy resolution results of the \textit{Xfluo} campaign are summarized in Table \ref{tab:Xfluo_G+C_FWHM}. We note that the energy resolution results obtained with \textit{Xfluo} and \textit{SOLEIL} data are marginally compatible at 3$\sigma$ level (for instance, FWHM is measured at 1.486 keV to be ~164 $\pm$ 3 eV with the \textit{Xfluo} source, while with the \textit{SOLEIL} data we measure at 1.550 keV a FWHM of ~151 $\pm$ 1 eV). In both cases, the increase of the FWHM up to 234\% at 1.5~keV is below the requirement of a 250\% increase (i.e. below the 200~eV requirement).
On the non-irradiated side with the \textit{Xfluo} source, we measure an energy resolution of ~83 $\pm$ 1~eV for the $Al$ $K\alpha$ line (at $\approx$ 1.5~keV), slightly higher than the measurement before irradiation measured on the flight model \cite{Benjamin}, of $\approx$ 70~eV, but this is expected, since the testing conditions are not completely equivalent (stand alone focal plane vs fully integrated camera).

\begin{figure*}[]
    \centering
    \includegraphics[scale = 0.5]{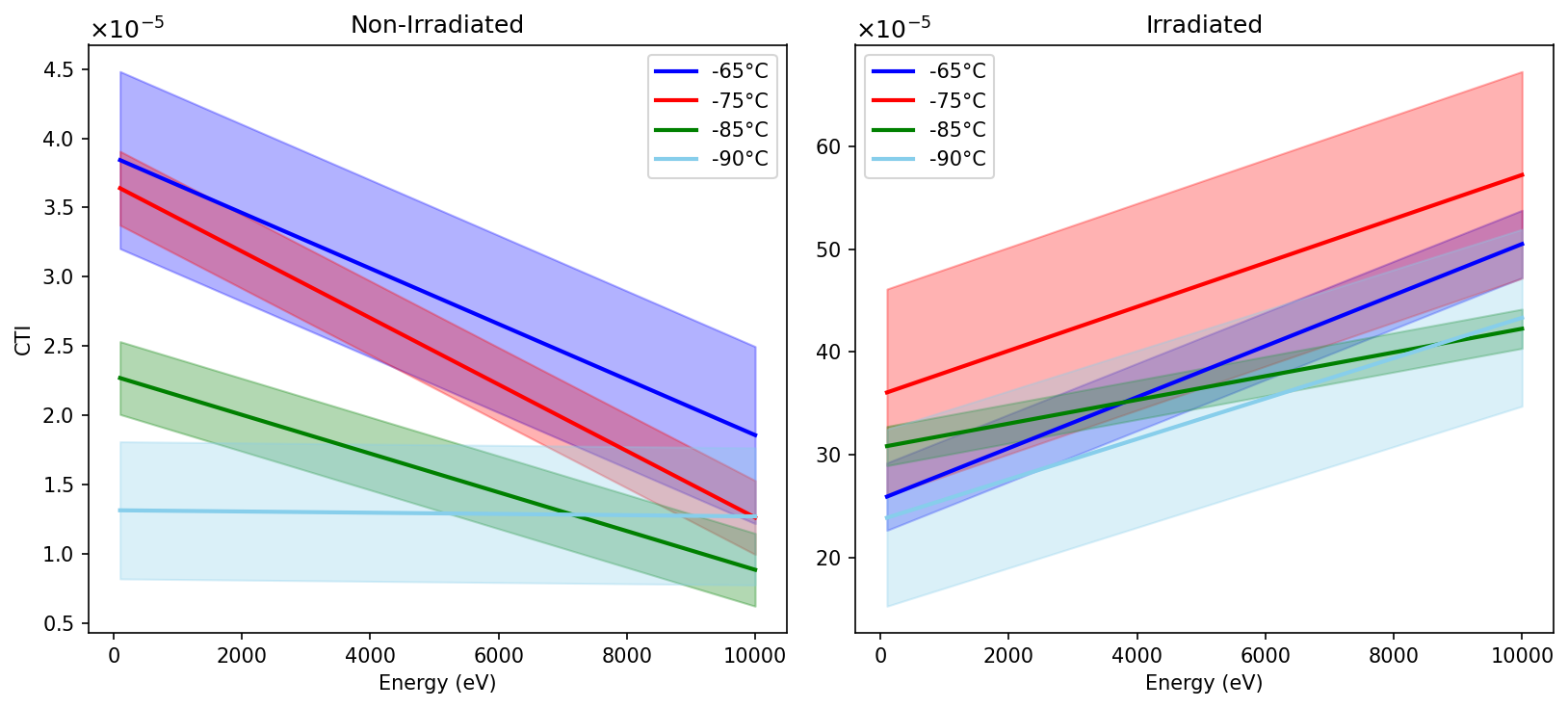}
    \caption{CTI results from \textit{Xfluo} data as a function of energy for the irradiated (right) and non-irradiated (left) detectors at multiple temperatures. Different colors represent the various detector temperatures. Data points have been omitted for clarity. The shaded area represent the standard error of the estimated linear fit.} 
    \label{fig:CTIvsE_genX}
\end{figure*}

\section{Discussion}
\label{sect:discussions}

\subsection{Summary of the results}
Some effect were expected and recovered after irradiation:
\begin{itemize}
    \item CTI values on the non-irradiated part are consistent with data before irradiation and with the flight model.
    \item The main parameter affected by the temperature is the dark current. A temperature increase engenders a dark current increase and thus a dark noise increase. 
    This effect is not visible before irradiation because dominated by the electronic noise. 
    \item Charge trapping becomes significant as seen in the CTI increase, manifested by the left tailing of the X-ray lines, as well as the apparent detector gain decrease (in ADU/eV) due to the contribution of the frame store: when the CTI in the frame store becomes significant, less charges are seen in the detector (ADUs) to reconstruct the same incoming photon energy, i.e. the gain is smaller.
\end{itemize}

Some effects of irradiation however, were not expected, such as:
\begin{itemize}
    \item Detector gain dependence with temperature, more pronounced than the CTI, e.g. a detector gain decrease of 16\% from -90$^\circ$C to -65$^\circ$C, whereas no clear tendency of the CTI with temperature can be drawn. 
    \item The CTI increase with energy.
\end{itemize}

The possible origin of these unexpected effects will be discussed hereafter.

\subsection{CTI and detector gain dependency with temperature}
As shown in Figure \ref{fig:CTIvsE_genX}, the CTI is dependent on the detector operating temperature. This effect is difficult to characterize in our analysis, because as discussed above, the CTI also greatly depends on the energy (i.e. on the trap types and their cross section). What we can conclude from our analysis is that, since no prominent tendency of the CTI as a function of temperature can be drawn, the proton irradiation - at least in the image area - does not create traps type dependent of a characteristic temperature within the nominal operating range.

The gain change is not a change in the gain properties of the integrated JFETs (radiation hard detectors) or the CAMEX analog response (low total ionsing dose). As previously mentioned, our interpretation of the gain shift is a change in the CTI of the frame store. CTI is associated to phenomena of electron capture, detrapping, thermal generation, whose time constants depend on the temperature in a semiconductor.

The CTI in the frame store (i.e. the gain loss) seems to be very dependent on the temperature, whereas the CTI in the image area shows no prominent tendency with temperature. We know that the defect types are the same in the two areas since they are irradiated in the same way. However the timings involved are very different: the time between two transfers is ~30~$\mu$s in the frame store and 1~$\mu$s in the image area. A final interpretation would require to update our theoretical model of CTI \citep{francesco_these} with the defect types, the time constants and energy levels. Those data are very difficult to obtain by our group and would require further physico-chemical analysis.

\subsection{CTI dependency with energy}

After irradiation, the CTI increases with energy with a seemingly linear trend. This is at odds with what has been measured for the non-irradiated flight detector \citep{Benjamin}, for which the CTI is decreasing with energy (see Figure \ref{fig:CTIvsE_genX}). Quantitatively we measure from 300~eV to 1850~eV an increase of CTI by a factor of $\sim$~1.4, whereas on the non-irradiated detector the CTI has E$^{-0.42}$ energy dependence \citep{Benjamin}, which in the same range gives a decrease by a factor of $\sim$~2.1. The different behaviour of the CTI with energy between the non-irradiated and irradiated detector may be explained by the fact that we are not dominated by the same traps before and after the irradiation. However, it is worth pointing out that the currently available theoretical model of the CTI \citep{francesco_these} cannot reproduce this tendency, no matter the properties of the defect types. One possible explanation would be that the modeling of the volume where the charge is confined is assumed independent of the energy while it is probably not. Here again, our group has no access to further information to refine the model.

Additionally, after irradiation, significantly different values of CTI and in particular CTI fitting laws have been obtained depending on the X-ray source. 
The difference in CTI values obtained between different calibration campaigns can qualitatively be explained by intrinsic differences of the sources. Table \ref{tab:sources_comparison} resumes the main sources characteristics for comparison.

\begin{table}[ht]
    \begin{center}
    \caption{Comparison of the sources based on basic illumination properties and event statistics. The multiplicity is the number of hit pixel for a single event. } 
    \scalebox{0.7}{
    \begin{tabular}{c|c|c|c}
    \rule[-1ex]{0pt}{3.5ex} & \textit{SOLEIL} & \textit{Xfluo}  &  \textit{${}^{55}$Fe} \\ 
    \hline
    \rule[-1ex]{0pt}{3.5ex} Flux by second (photon/s) & 150 & 9000 & 1000 \\
    \hline
    \rule[-1ex]{0pt}{3.5ex} Flux by pixel (photon/pixel/s) & \textit{on patch:} $1.8.10^{-1}$ & $1.4.10^{-1}$ & $1.6.10^{-2}$ \\ 
    \rule[-1ex]{0pt}{3.5ex}  & \textit{full matrix:} $3.2.10^{-2}$ &  & \\
    \end{tabular}}
    \label{tab:sources_comparison}
    \end{center}
\end{table}

\begin{figure}[]
    \centering
    \includegraphics[scale = 0.4]{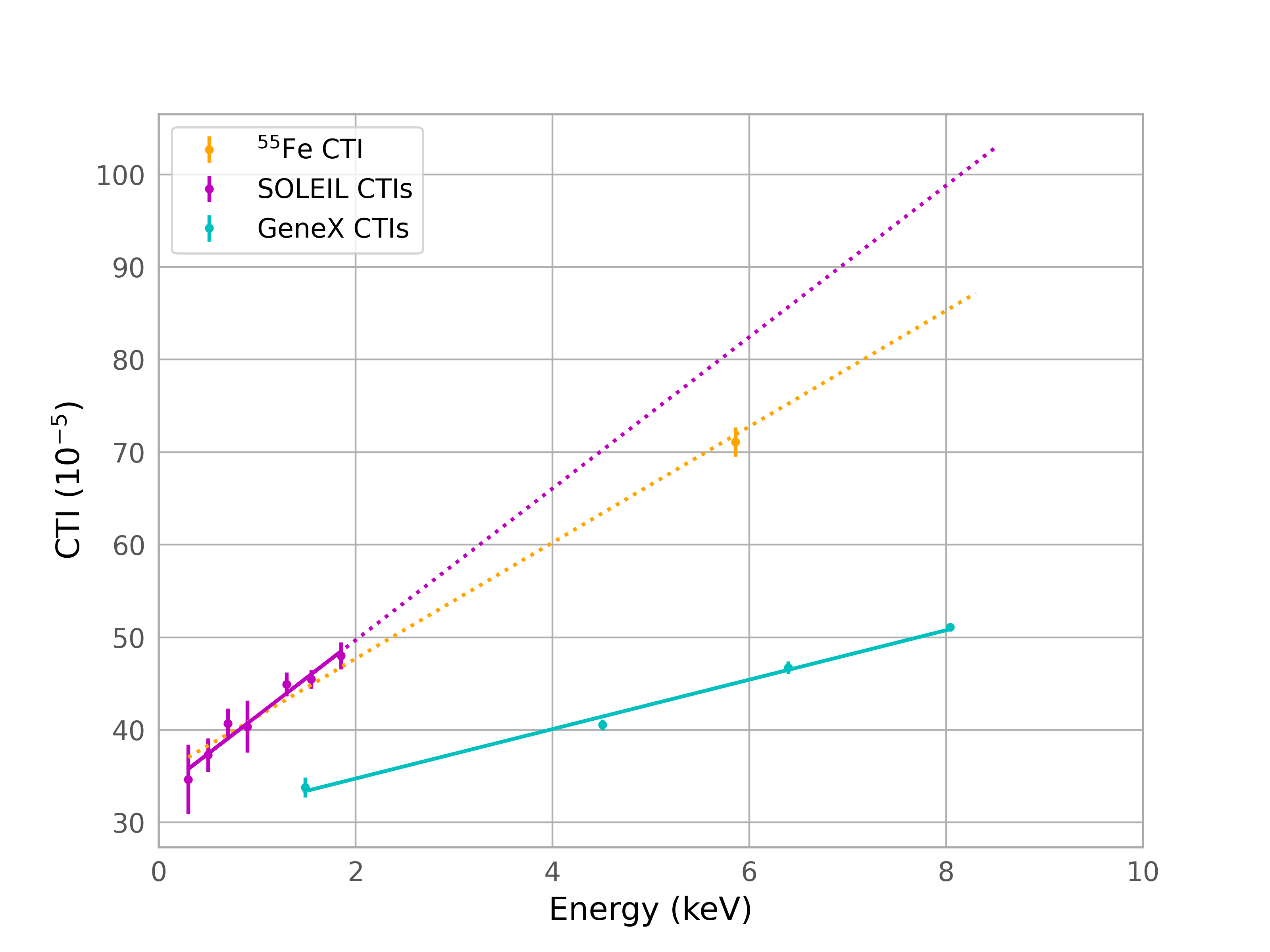}
    \caption{Plot summarizing all CTI measurements obtained over the different campaigns (with the irradiated detector, at -65$^\circ$) - in purple for \textit{SOLEIL}, in cyan for \textit{Xfluo}, and in orange for \textit{SOLEIL} including the $^{55}$Fe measurement - with their corresponding linear evolution with energy laws (filled lines). The dashed purple and orange lines represent the extrapolated \textit{SOLEIL} laws, respectively including and not including the $^{55}$Fe point, over MXT energy range. } 
    \label{fig:all_ctis}
\end{figure}

For a bright and uniform source such as \textit{Xfluo}, electron traps are more likely to be filled by existing charges, which reduces the probability of trapping during the charge transfer \citep{Meidinger_XMM}. The CTI values for a given energy are thus lower with a \textit{Xfluo}-like source, as is indeed measured Figure \ref{fig:all_ctis}, displaying the CTI results of MXT irradiated detector with each source, as well as the measurement of the internal $^{55}$Fe calibration source. We note that the CTI with the $^{55}$Fe calibration source illuminating the irradiated detector is $(71.5 \pm 2.8 )~10^{-5}$, being more in line with the {\it SOLEIL} results. This is because although the $^{55}$Fe calibration source illuminates the entire detector, like \textit{Xfluo}, its flux is significantly lower (see Table \ref{tab:sources_comparison}, reducing the trap filling effect. 
In that sense, the \textit{Xfluo} source is useful to cover MXT energy range, but it does not realistically reproduce the illumination expected from MXT intended targets. In flight, the X-rays will be rather concentrated in specific areas of the detector due to the focusing of the optics and the expected fluxes are at most a few hundreds of counts per second over the entire detector. Hence the conditions will be more similar to the {\it SOLEIL} experiment, where contrary to \textit{Xfluo}, the flux is higher on the beam patch, but is significantly lower averaged over the whole detector matrix (see Table \ref{tab:sources_comparison}). As a consequence, due to both its energy coverage where the optics is the most effective and its overall flux closer to astrophysical sources, the \textit{SOLEIL} data is more appropriate to predict the detector in-flight performance. This will be our preferred model against which our prediction will be tested using in-flight calibration data.

\section{Conclusion}

In this work, we presented the various characterization campaigns performed on the irradiated spare model of MXT detector to anticipate the evolution of its spectral performance over the entire SVOM mission. The proton irradiation, that we reproduced on a spare model of MXT solely for this work, creates defects in the silicon lattice by displacement damage. This induces an increase in dark current, as seen in the dark noise increase, which is susceptible to negatively impact the detection capabilities of MXT as it increases the lowest energy detectable. Additionally, the defects in the lattice act as traps for the charges, increasing the trapping probability, as seen in the CTI increase, which causes shifts of the spectral lines to lower energies, degrades the spectral resolution, and disrupts the symmetry of the spectral lines. We chose to irradiate the spare model of the MXT camera with a 6$\times$10$^9$ protons.cm$^{-2}$ fluence, accounting for the duration of the nominal mission (3 years). The CTI is the main contributor to the degradation of spectral performances due to irradiation. We showed that the CTI increases with energy and that the correction shall be a function of the energy. It must be corrected to evaluate performance, especially at high energies where it largely dominates. The establishment of a CTI law with energy is thus decisive, yet complex because this evolution is misunderstood. The behaviour of the CTI with energy can not be reproduced by our modelling, and causes significant interpretation issues. There is a lack of representation in literature concerning this CTI evolution. Further work on the CTI evolution with energy would be essential to fully understand this issue. Consequently, we conclude that the calibration source present aboard, due to its limited number of spectral lines, will not suffice to monitor the evolution of MXT camera response throughout the life span of the mission; relatively frequent visit to sky calibration sources, such as Supernova remnants, will be essential. 
However, our analysis on the irradiated spare model demonstrates that, when corrected for CTI, MXT remains compliant with the instrument requirements. First regarding the energy range, despite an leakage current increase at the nominal detector temperature that can not be corrected, 200~eV energy photons can still be detected. Regarding the energy resolution, the increase from the beginning of life to the end of life from about 70~eV to about 150~eV at 1.5~keV is well below the 200~eV EoL requirement. Hence, the MXT camera will operate to its required performances over the whole span of the nominal mission, with some margin. The different calibration campaigns presented here will be of great support during MXT in flight operations in order to better understand the future behaviour of the detector.

\subsection* {Acknowledgments}
The MXT has been developed by CNES in collaboration with CEA, University of Leicester, MPE, and IJCLab. 
We acknowledge the excellent support of the \textit{SOLEIL} team (P. Mercère, P. Da Silva) and the ARRONAX team (C. Koumeir, F. Haddad) during the test campaigns. We thank N. Meidinger and O. Limousin for useful discussions during the preparation of the manuscript.\\


\appendix

\section*{Appendix}

\subsection*{Appendix A: \textit{Crystal ball} fit for the measurement of the energy resolution from \textit{SOLEIL} data}
\label{sec:appendix:soleil_fit}
\begin{figure}[h!]
    \centering
    \includegraphics[scale = 0.4]{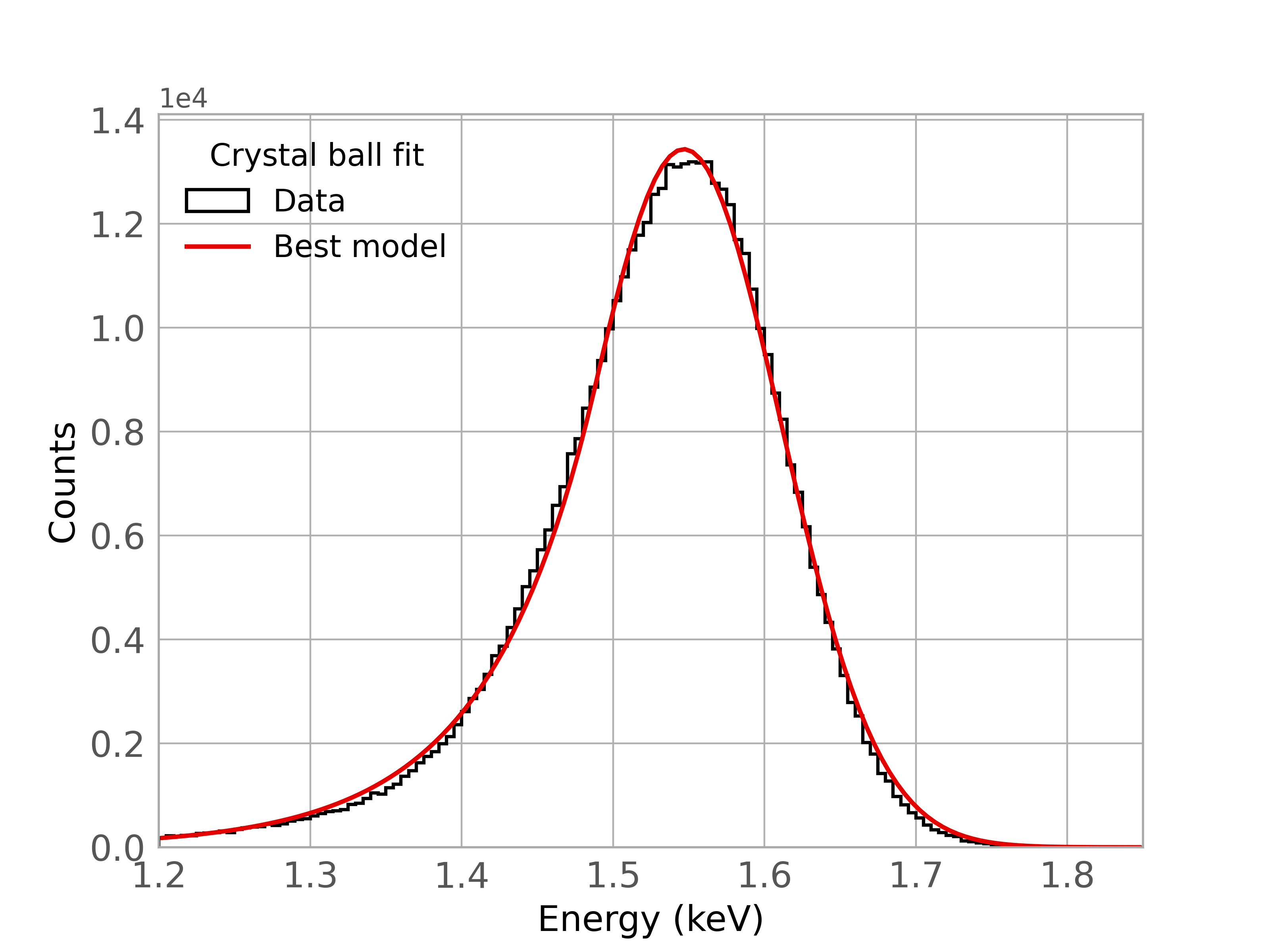}
    \caption{Example of a \textit{Crystal ball} fit of \textit{SOLEIL} data at 1550~eV from the irradiated detector to measure the end of life energy resolution. }
    \label{fig:fit_SOLEIL}
\end{figure}

For \textit{SOLEIL} data, to account for the energy loss, due to the CTI, rendering the spectral lines asymmetric, we opted for the \textit{Crystal ball} function to measure the energy resolution. Often, the lower energy tail is not closely fitted, but the width at half maximum of the spectral line is well reproduced, making the \textit{Crystal ball} function a suitable choice for the determination of the EoL energy resolution, see Figure \ref{fig:fit_SOLEIL}. 

\subsection*{Appendix B: Gaussian + Quadratic fit for the measurement of the energy resolution from \textit{Xfluo} data}
\label{sec:appendix:Xfluo_fit}

\begin{figure}[h!]
    \centering
    \includegraphics[scale = 0.4]{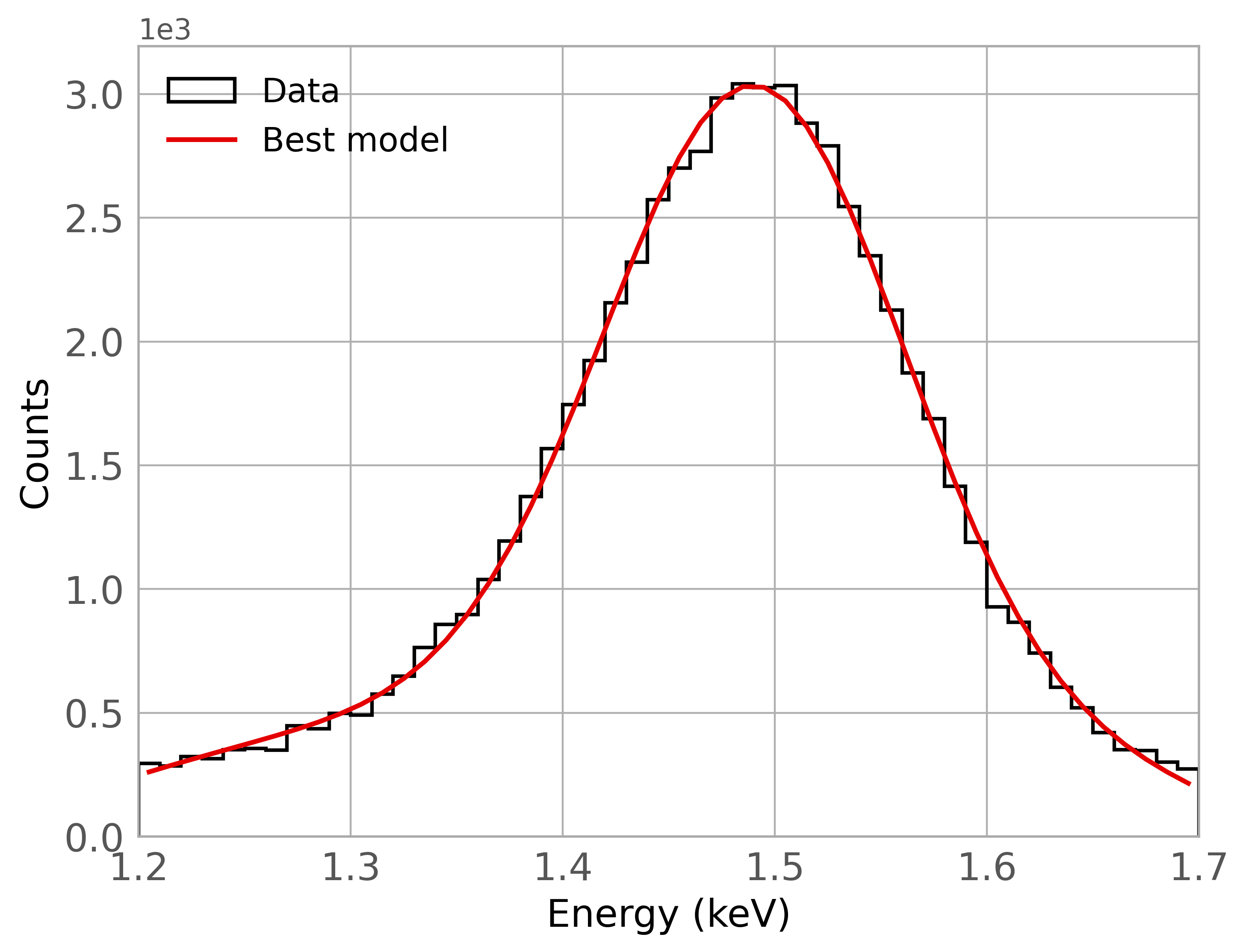}
    \caption{Example of a Gaussian + Quadratic fit of \textit{Xfluo} data for the $Al$ $K\alpha$ line (at 1.5~keV) from the irradiated detector to measure the end of life energy resolution. }
    \label{fig:fit_Xfluo}
\end{figure}

For \textit{Xfluo} data, a Gaussian fit with a quadratic component was able to provide a satisfactory fit, see Figure \ref{fig:fit_Xfluo}.

\bibliography{cas-refs}

\end{document}